\documentclass[twocolumn]{aastex631}
\usepackage{soul}
\usepackage{graphicx}
\usepackage{subfigure}

\shorttitle{}
\shortauthors{Yu et al.}
\graphicspath{{./}{figures/}}

\begin{document}

\title{Hilbert--Huang Transform analysis of quasi-periodic oscillations in MAXI J1820+070 }

\correspondingauthor{Qing-cui Bu}
\email{bu@astro.uni-tuebingen.de}

\author{Wei Yu}
\affiliation{Key Laboratory of Particle Astrophysics, Institute of High Energy Physics, Chinese Academy of Sciences, 19B Yuquan Road, Beijing 100049, China}
\affiliation{University of Chinese Academy of Sciences, Chinese Academy of Sciences, Beijing 100049, China}
\author{Qing-Cui Bu}
\affiliation{Institut f{\"u}r Astronomie und Astrophysik, Kepler Center for Astro and Particle Physics, Eberhard Karls Universit{\"a}t, 72076 T{\"u}bingen, Germany}

\author{Zi-Xu Yang}
\affiliation{Key Laboratory of Particle Astrophysics, Institute of High Energy Physics, Chinese Academy of Sciences, 19B Yuquan Road, Beijing 100049, China}
\affiliation{University of Chinese Academy of Sciences, Chinese Academy of Sciences, Beijing 100049, China}

\author{He-Xin Liu}
\affiliation{Key Laboratory of Particle Astrophysics, Institute of High Energy Physics, Chinese Academy of Sciences, 19B Yuquan Road, Beijing 100049, China}
\affiliation{University of Chinese Academy of Sciences, Chinese Academy of Sciences, Beijing 100049, China}

\author{Liang Zhang}
\affiliation{Key Laboratory of Particle Astrophysics, Institute of High Energy Physics, Chinese Academy of Sciences, 19B Yuquan Road, Beijing 100049, China}

\author{Yue Huang}
\affiliation{Key Laboratory of Particle Astrophysics, Institute of High Energy Physics, Chinese Academy of Sciences, 19B Yuquan Road, Beijing 100049, China}

\author{Deng-Ke Zhou}
\affiliation{Research Center for Intelligent Computing Platforms, Zhejiang Laboratory, Hangzhou 311100, China}

\author{Jin-Lu Qu}
\affiliation{Key Laboratory of Particle Astrophysics, Institute of High Energy Physics, Chinese Academy of Sciences, 19B Yuquan Road, Beijing 100049, China}

\author{Shuang-Nan Zhang}
\affiliation{Key Laboratory of Particle Astrophysics, Institute of High Energy Physics, Chinese Academy of Sciences, 19B Yuquan Road, Beijing 100049, China}
\affiliation{University of Chinese Academy of Sciences, Chinese Academy of Sciences, Beijing 100049, China}

\author{Shu Zhang}
\affiliation{Key Laboratory of Particle Astrophysics, Institute of High Energy Physics, Chinese Academy of Sciences, 19B Yuquan Road, Beijing 100049, China}

\author{Li-Ming Song}
\affiliation{Key Laboratory of Particle Astrophysics, Institute of High Energy Physics, Chinese Academy of Sciences, 19B Yuquan Road, Beijing 100049, China}
\affiliation{University of Chinese Academy of Sciences, Chinese Academy of Sciences, Beijing 100049, China}

\author{Shu-Mei Jia}
\affiliation{Key Laboratory of Particle Astrophysics, Institute of High Energy Physics, Chinese Academy of Sciences, 19B Yuquan Road, Beijing 100049, China}

\author{Xiang Ma}
\affiliation{Key Laboratory of Particle Astrophysics, Institute of High Energy Physics, Chinese Academy of Sciences, 19B Yuquan Road, Beijing 100049, China}

\author{Lian Tao}
\affiliation{Key Laboratory of Particle Astrophysics, Institute of High Energy Physics, Chinese Academy of Sciences, 19B Yuquan Road, Beijing 100049, China}

\author{Ming-Yu Ge}
\affiliation{Key Laboratory of Particle Astrophysics, Institute of High Energy Physics, Chinese Academy of Sciences, 19B Yuquan Road, Beijing 100049, China}

\author{Qing-Zhong Liu}
\affiliation{Purple Mountain Observatory, Chinese Academy of Sciences, Nanjing 210008, China}

\author{Jing-Zhi Yan}
\affiliation{Purple Mountain Observatory, Chinese Academy of Sciences, Nanjing 210008, China}

\begin{abstract}

We present time-frequency analysis, based on the Hilbert--Huang transform (HHT), of the evolution on the low-frequency quasi-periodic oscillations (LFQPOs) observed in the black hole X-ray binary MAXI J1820+070. Through the empirical mode decomposition (EMD) method, we decompose the light curve of the QPO component and measure its intrinsic phase lag between photons from different energy bands. We find that the QPO phase lag is negative (low energy photons lag behind high energy photons), meanwhile the absolute value of the lag increases with energy. By applying the Hilbert transform to the light curve of the QPO, we further extract the instantaneous frequency and amplitude of the QPO. Compared these results with those from the Fourier analysis, we find that the broadening of the QPO peak is mainly caused by the frequency modulation. Through further analysis, we find that these modulations could share a common physical origin with the broad-band noise, and can be well explained by the internal shock model of the jet.

\end{abstract}



\section{Introduction} \label{sec:intro}

\begin{figure*}
	\centering\includegraphics[width=\columnwidth]{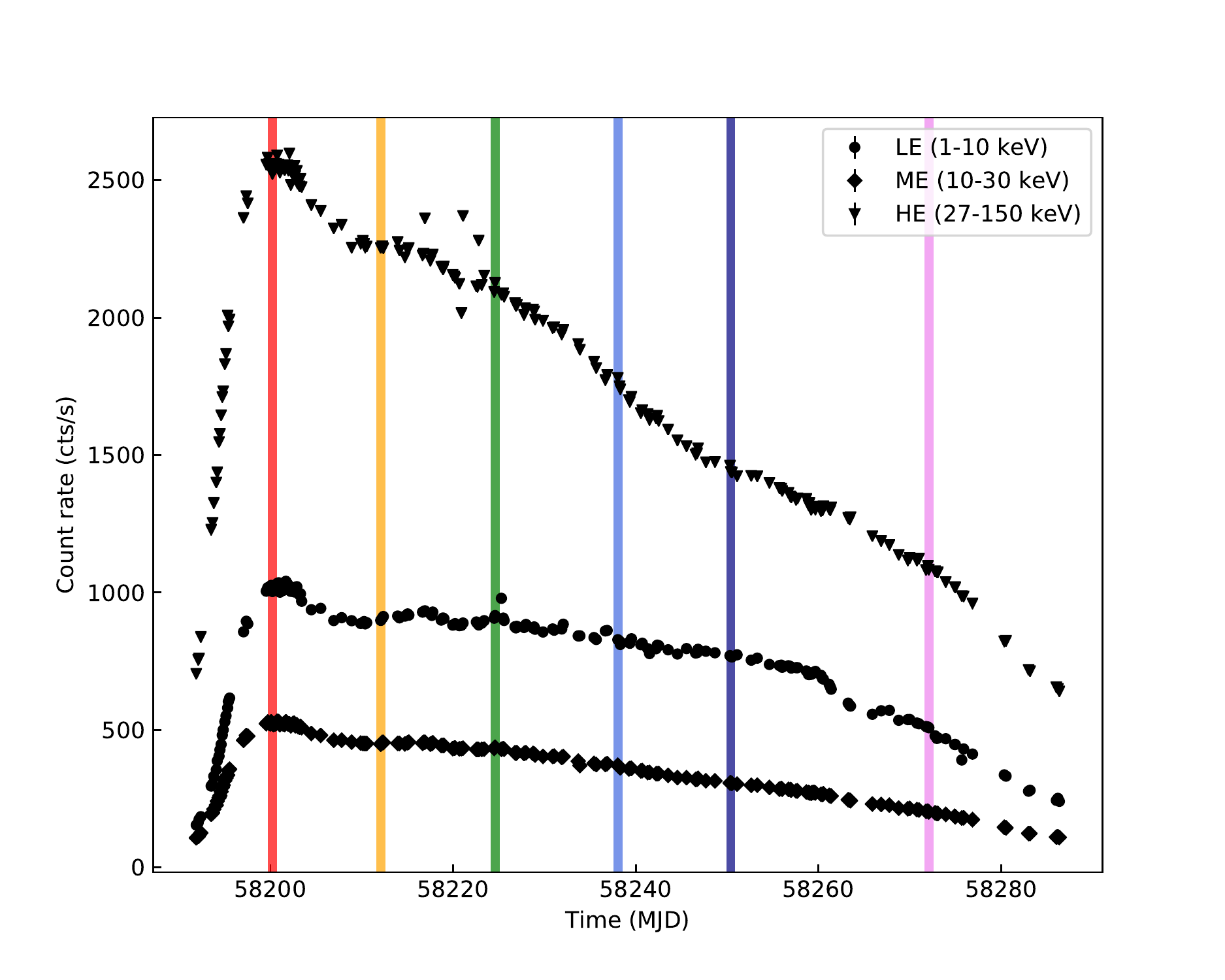}
	\centering\includegraphics[width=\columnwidth]{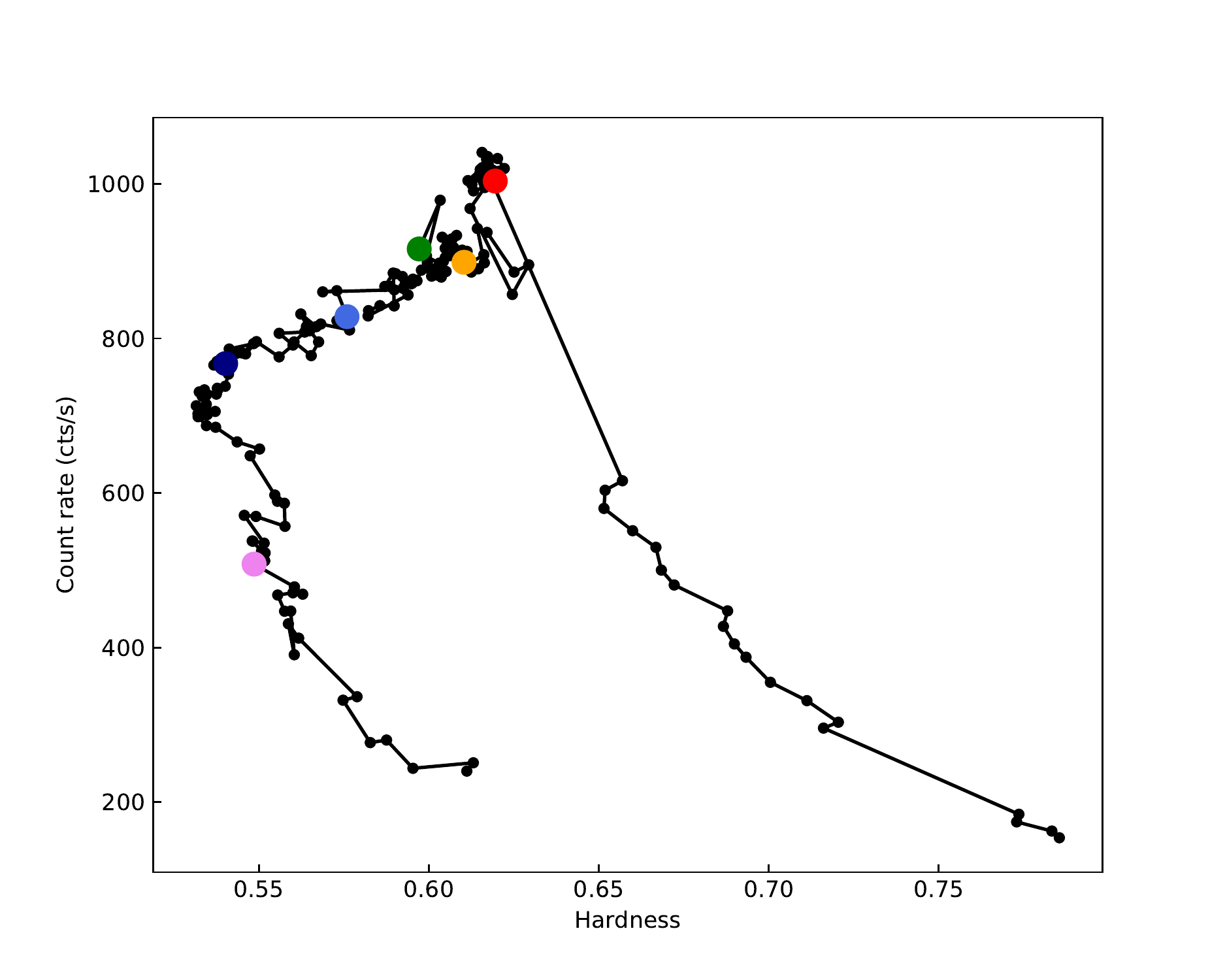}
    \caption{Left: \emph{Insight}-HXMT lightcurves of MAXI J1820+070 from HE (27--150 keV), ME (10--30 keV) and LE (1--10 keV), extracted from the rising hard state during its 2018 outburst from MJD 58191 to 58286. Right: The corresponding hardness-intensity diagram, with the intensity defined as the total count rate of 1--10 keV and hardness defined as the ratio between the count rates from the hard band (3--10 keV) and the soft band (1--3 keV). The six epochs selected for our study are highlighted in different colours.}
    \label{fig1}
\end{figure*}

\begin{figure*}
	\centering\includegraphics[width=2\columnwidth]{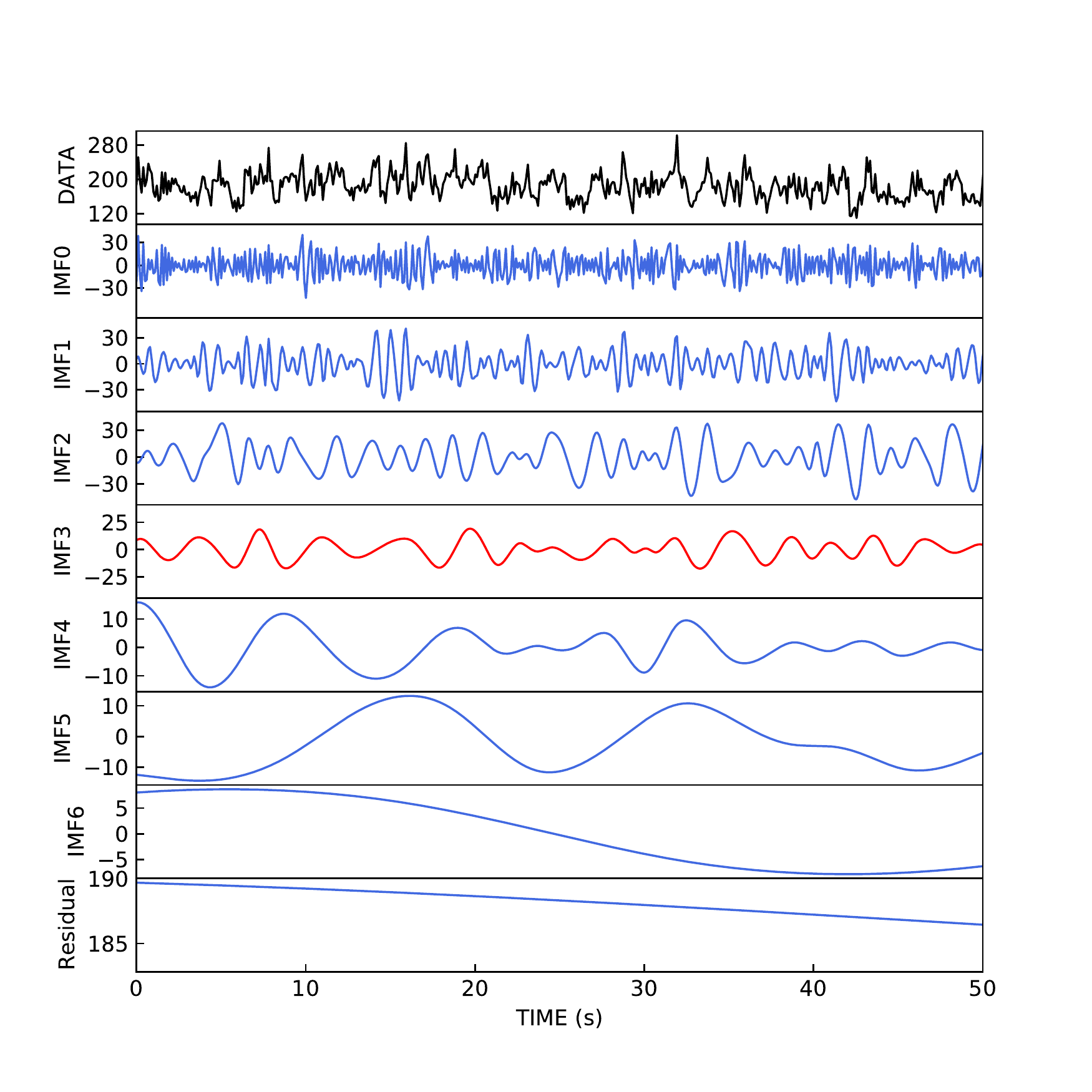}
    \caption{Representative example of a 50-s-long lightcurve from 27--150 keV and its corresponding IMFs. From the top to bottom: the original light curve (DATA); the high-frequency noises from the summation of IMF0 to IMF2; the LFQPO (IMF3); the low frequency noise from the summation of IMF4 to the residual.}
    \label{fig2}
\end{figure*}

Low-frequency quasi-periodic oscillations (LFQPOs) with frequencies ranging from a few millihertz to $\sim$30 Hz have been found in most transient black hole X-ray binaries \citep{motta2015geometrical,motta2016quasi,belloni2010states,ingram2019review}. These oscillations are appropriately named to reflect their finite-width peaks, usually described by multi-Lorentzian components \citep{belloni2002unified,rao2010low}, in the Fourier power spectra of their X-ray light curves. LFQPOs are believed to originate from the inner part of the accretion flow, however, the physical mechanism is still not well understood. Theories about the origin of LFQPOs can be generally divided into two broad categories: intrinsic models associated with wave modes of the accretion flow \citep{tagger1999accretion,cabanac2010variability} and geometric effects such as relativistic precession of the inner hot flow or the jet due to a misalignment of the black hole spin axis and the binary orbital axis \citep{stella1997lense,stella1999correlations,ingram2009low}.

The LFQPO’s broad peak implies that the corresponding X-ray light curve is not strictly periodic and could be caused by a modulation with varying frequency or amplitude \citep{ingram2019review}. Since in the framework of Fourier analysis, the Fourier frequencies are defined as constant over the entire time, it doesn't give information of the variability of frequencies. Therefore, time-frequency analysis techniques are required to study the origin of the QPO peak broadening. One possible method to investigate the variation of QPO period is the Hilbert--Huang transform (HHT) proposed by \citet{huang1998empirical}. The HHT is a powerful tool for analyzing phenomena with non-stationary periodicity and has been successfully applied in astronomical research, such as for the QPO in the active galactic nucleus RE J1034 + 396 \citep{hu2014tracking} and the $\sim$4 Hz QPO observed in the black hole X-ray binary XTE J1550-564 \citep{su2015characterizing}. The HHT is used to decompose a non-stationary signal into basis components and transform these components into instantaneous frequencies and amplitudes. The basis components are not based on any strictly mathematical form and are derived by the signal itself. In contrast, the Fourier and wavelet analysis decompose a signal based on trigonometric and wavelet functions, respectively. The instantaneous frequency, which is different from that in the Fourier analysis, is defined as the time derivative of the phase function. Therefore, the Hilbert spectrum can provide detailed information in both time and frequency domains. It is worth mentioning that the HHT method is based on the assumption that the signals are additive in the time domain. This means that the original signal can be expressed as the sum of basis components. Thus, HHT cannot be applied to multiplied or convoluted signals.

The LFQPO phase lag has been commonly detected in black hole X-ray binaries and has provided insights into the geometry and the radiation processes of the accretion flow. However, the phase lag directly measured at the QPO frequency using the lag-frequency spectrum is not the intrinsic lag because of the interference brought from the underneath strong band-noise \citep{ma2021discovery,zhou2022determination}. By using the HHT method, we are able to extract the independent light curve of QPO, which further allows us to measure the intrinsic phase lag of the QPO. This method has been applied to GRS 1915+105 to study the phase lag of LFQPOs \citep{van2016probing}, and makes great sense to probe the origin of the QPO. 

MAXI J1820+070 is a low-mass BHXB, discovered by the Monitor of All-sky X-ray Image (MAXI) on 11 March 2018. Follow-up observations were made by other X-ray telescopes, e.g., \emph{Swift}/BAT, \emph{INTEGRAL}, \emph{NuSTAR}, and \emph{NICER}. QPOs have been observed in MAXI J1820+070 in multiple wavebands, from optical \citep{yu2018further,yu2018detection,zampieri2018low,fiori2018other} to hard X-ray \citep{mereminskiy2018low}. \emph{Insight}-HXMT carried out a Target of Opportunity (ToO) observation three days after its discovery, and monitored the whole outburst from 2018-03-14 (MJD 58191) to 2018-10-21 (MJD 58412). The high statistics and the broad energy coverage (1-250 keV) of \emph{Insight}-HXMT allow us to perform detailed timing analysis of the broadband variability, especially at high energy.

In this paper, we present the HHT-based analysis of the time-frequency properties of the LFQPOs detected in MAXI J1820+070. Through the HHT method, we measured the intrinsic phase lag of the QPO and explored the origin of its peak broadening. In Section 2, we briefly describe the \emph{Insight}-HXMT observation and the data reduction. In Section 3, we explain how to use the HHT for adaptive decomposition of the oscillatory component from the X-ray light curve, and present how to obtain the instantaneous frequency and amplitude of the QPO. The result and discussion are given in Section 4. At last, we end with our conclusion in Section 5.

\begin{figure}
	\centering\includegraphics[width=\columnwidth]{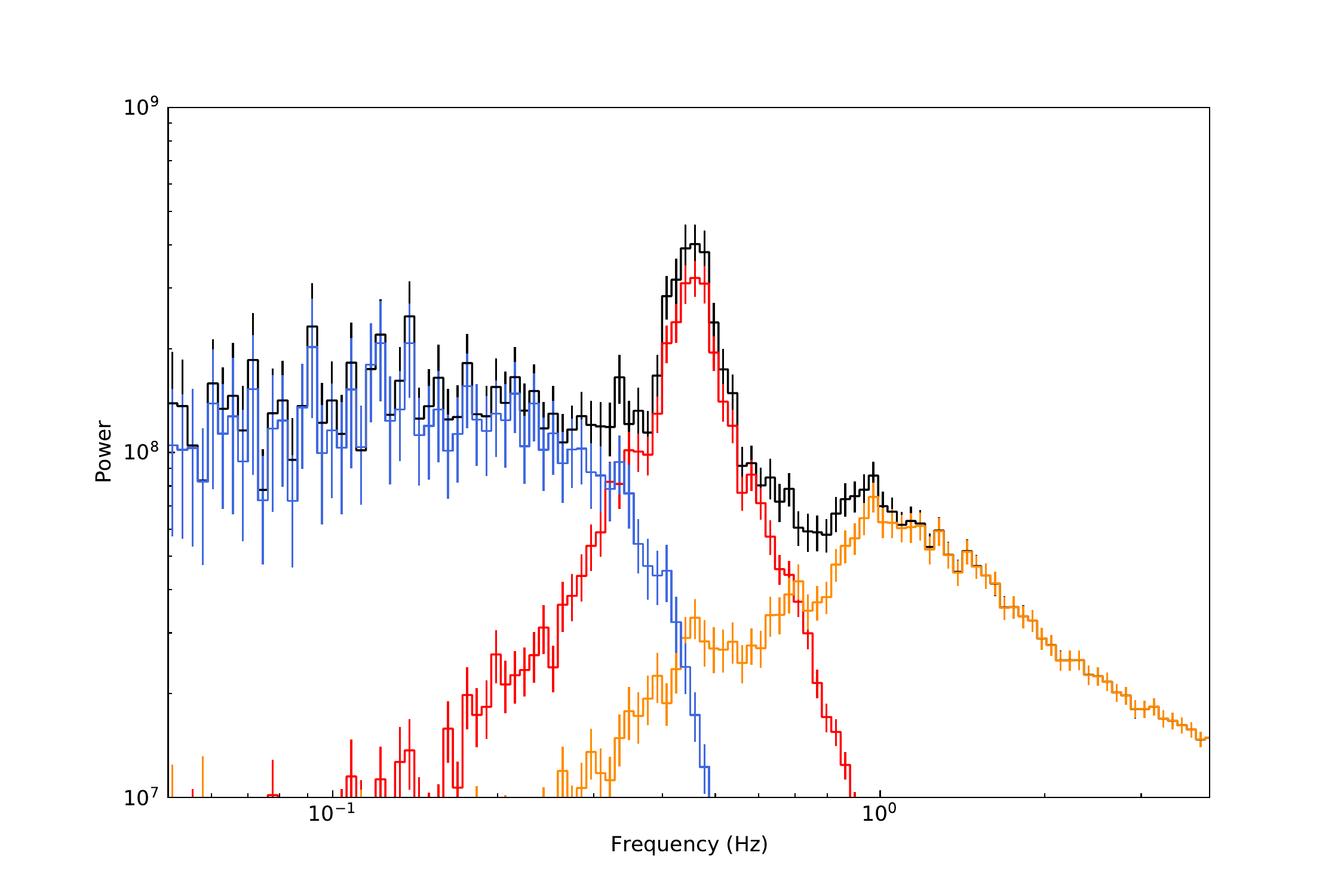}
    \caption{Fourier power spectra produced from the IMF0-2 (orange), IMF3 (red), IMF4-Residual (blue) and the original light curve (black).}
    \label{fig3}
\end{figure}

\begin{figure*}
	\centering\includegraphics[width=\columnwidth]{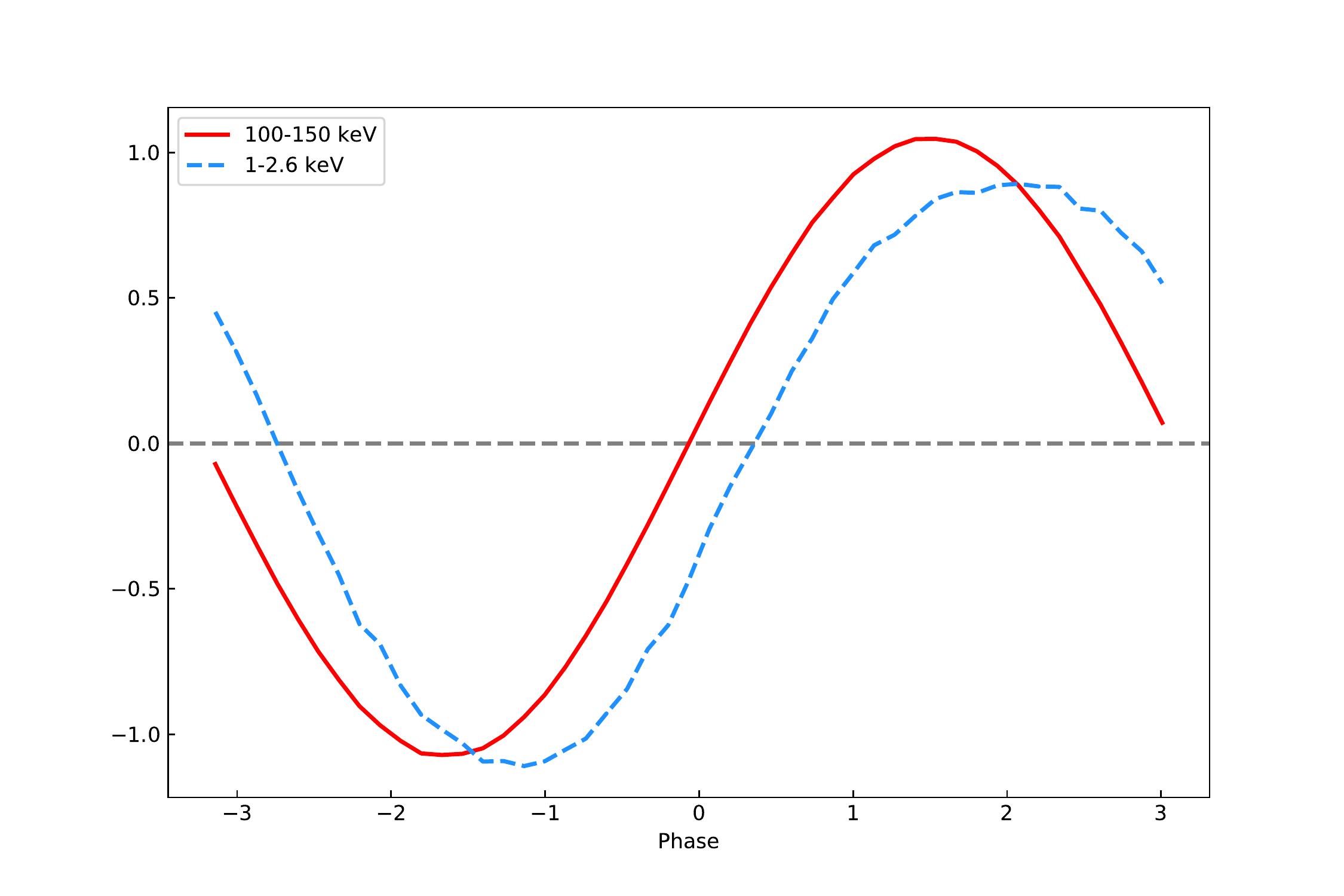}
	\centering\includegraphics[width=\columnwidth]{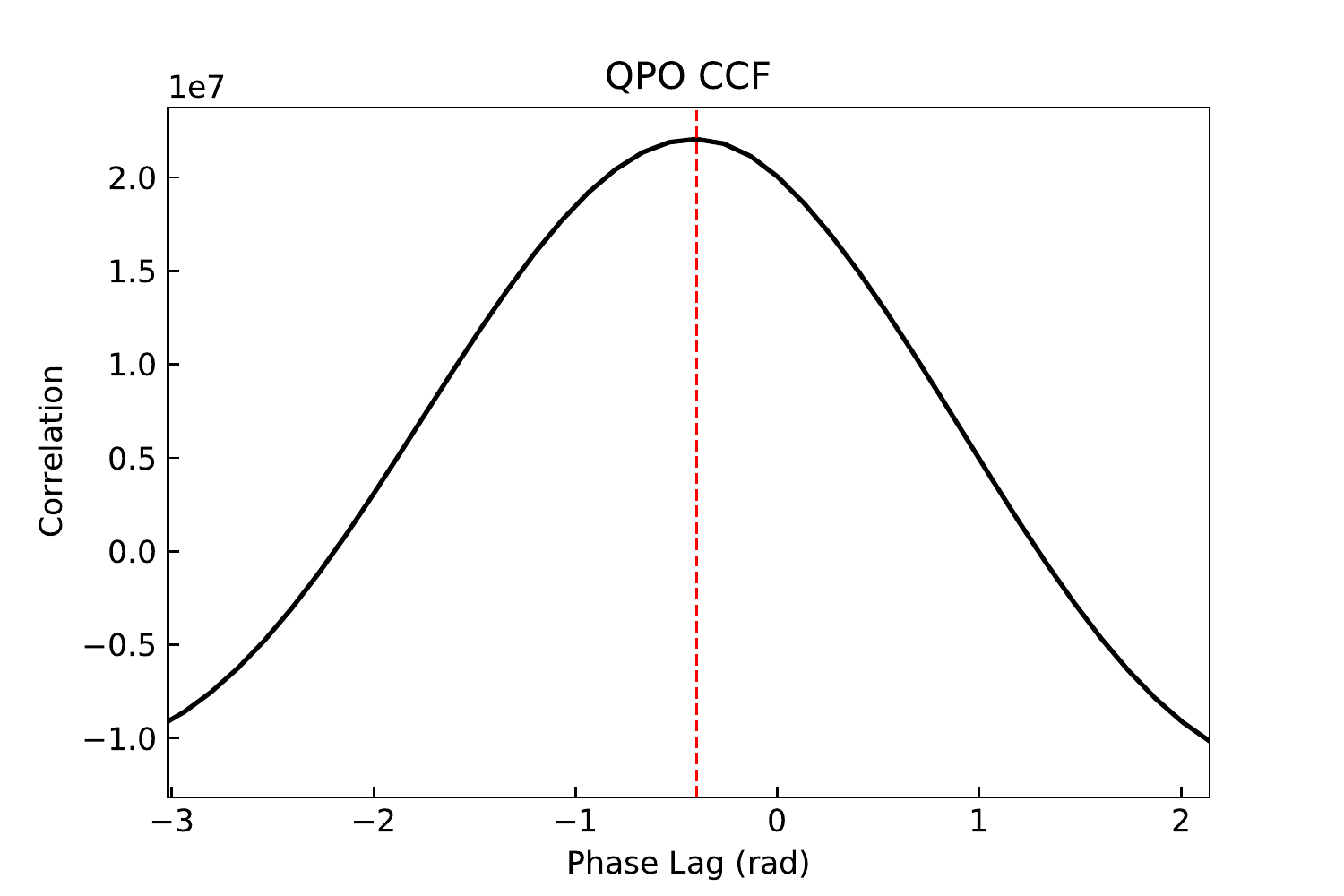}
    \caption{Left:The superimposed lightcurves from the energy band 1--2.6 keV (blue) and 100--150 keV (red). Right: cross-correlation function of IMF3 calculated between the 1--2.6 keV and 100--150 keV energy bands.}
    \label{fig4}
\end{figure*}

\begin{figure*}
	\centering\includegraphics[width=\columnwidth]{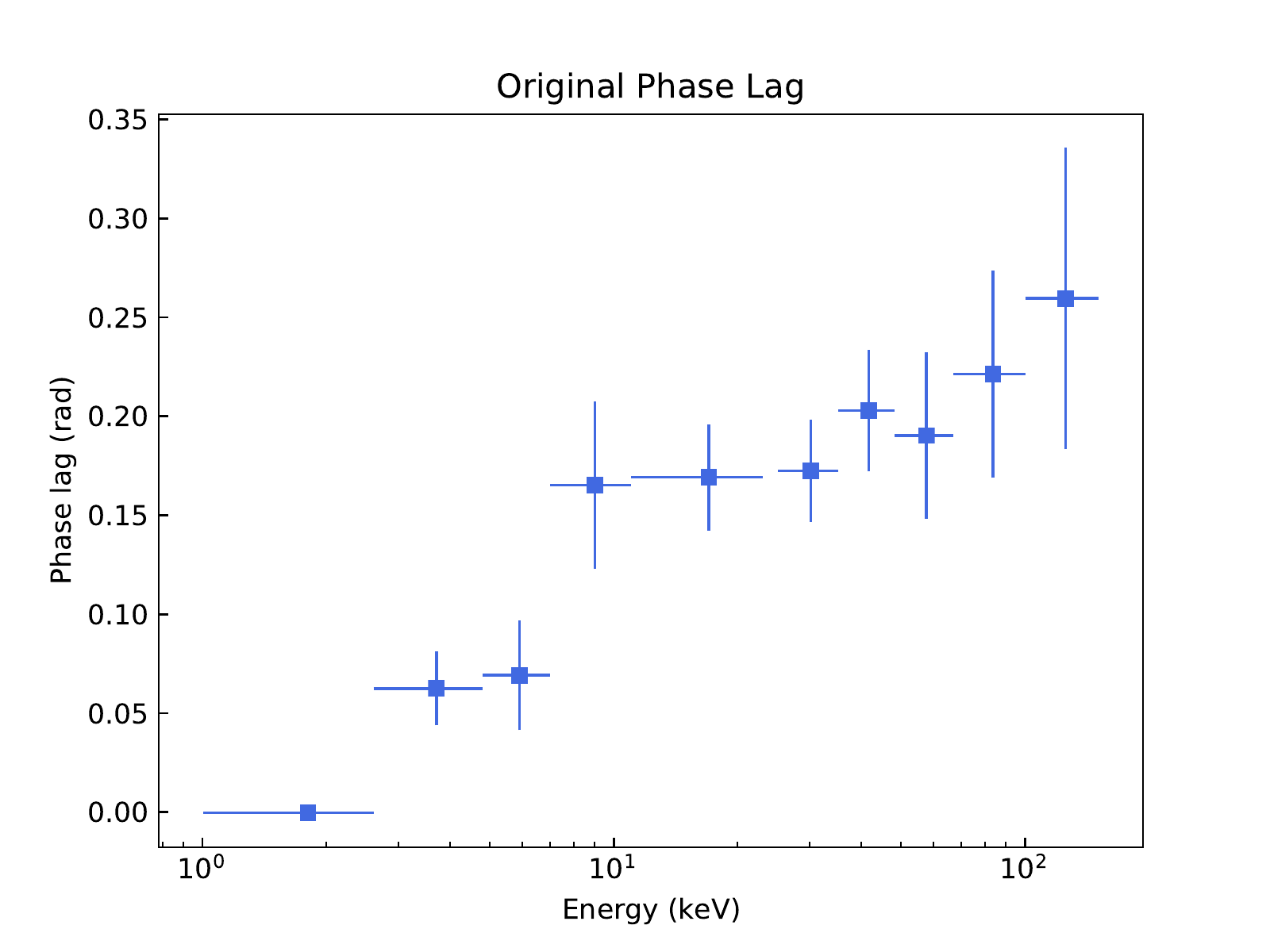}
	\centering\includegraphics[width=\columnwidth]{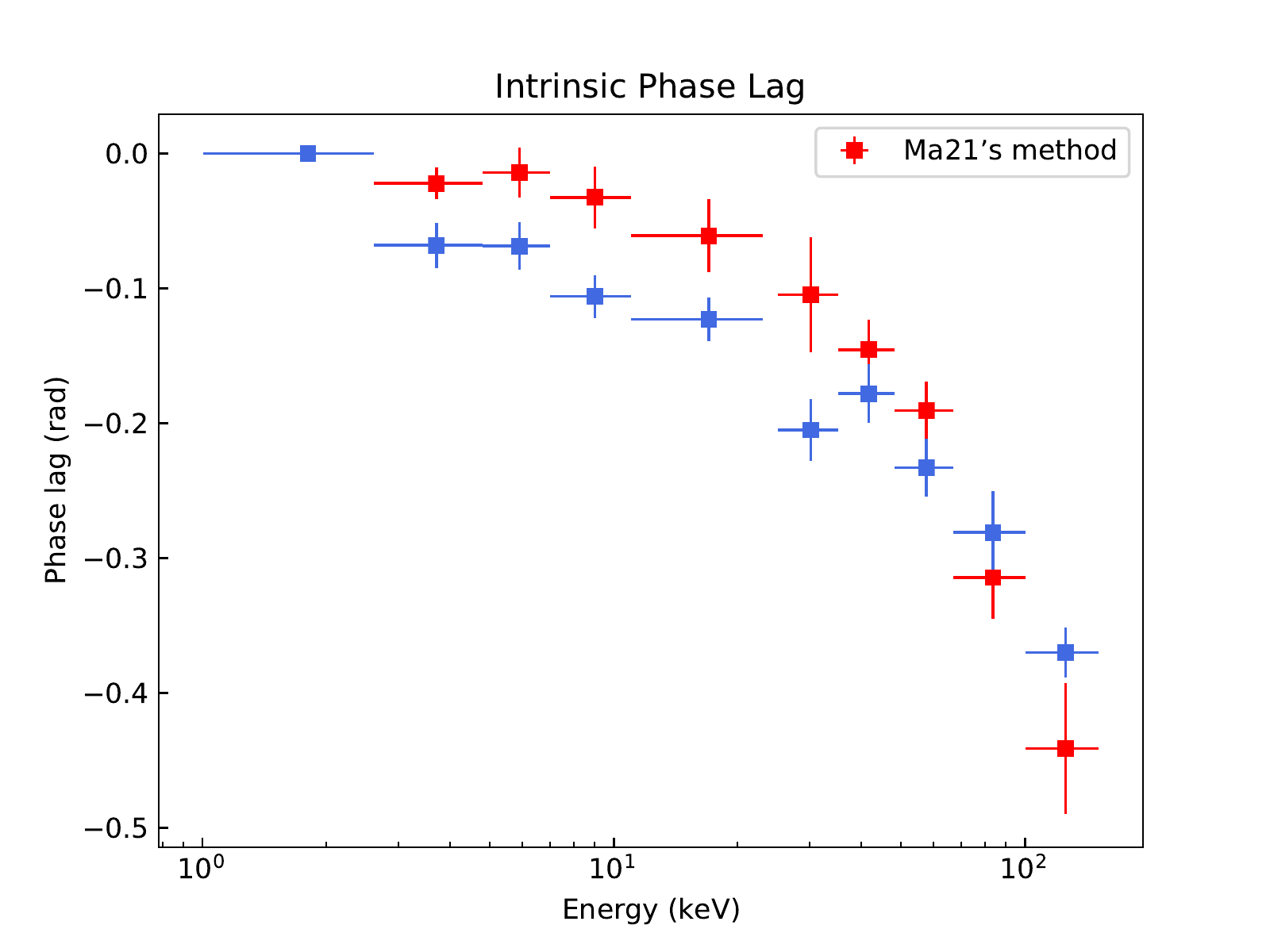}
    \caption{The LFQPO phase lags as a function of photon energy. The reference energy band is 1--2.6 keV. Left: the original phase lag. Right: the intrinsic phase lag. }
    \label{fig5}
\end{figure*}

\begin{figure*}
	\centering\includegraphics[width=2\columnwidth]{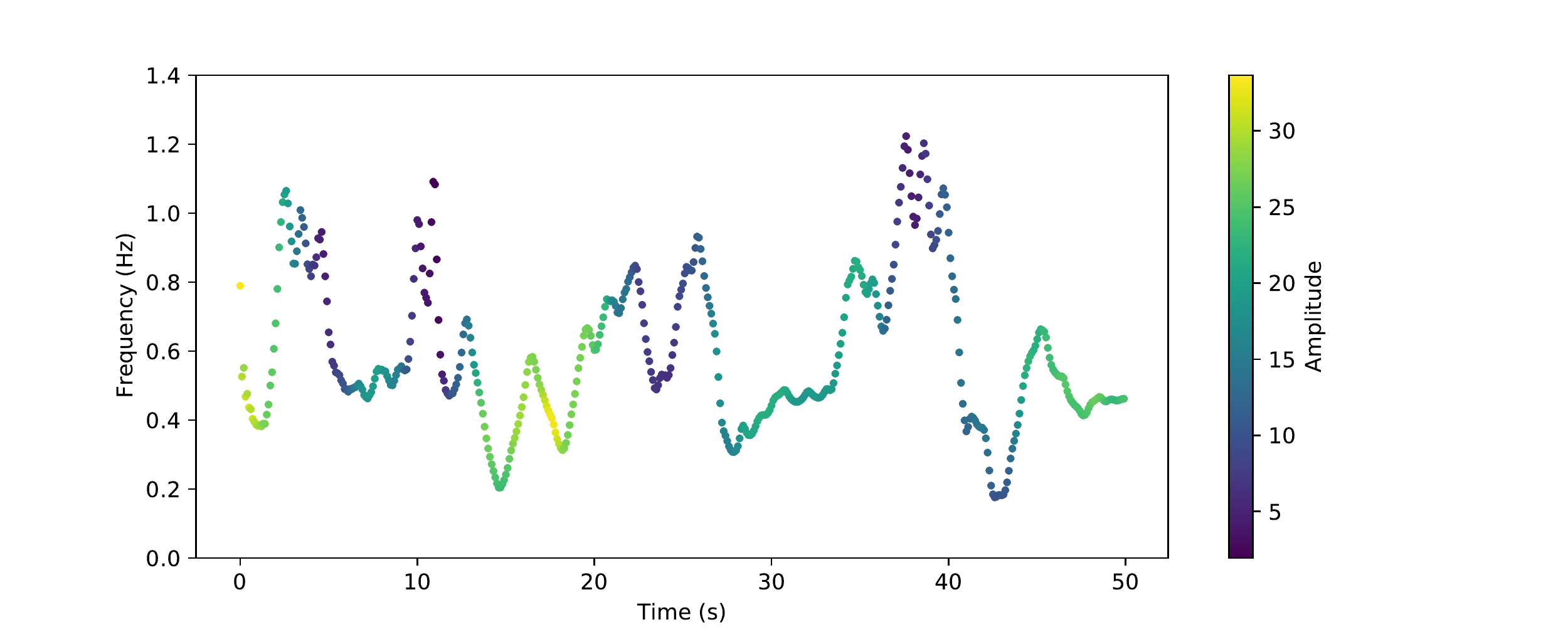}
    \caption{Hilbert spectrum of the LFQPO from MAXI J1820+070. The color on the z-axis represents the QPO amplitude.}
    \label{fig6}
\end{figure*}

\section{DATA REDUCTION} \label{sec2}

In this work, we use data observed with \emph{Insight}-HXMT between March 14th and June 17th, 2018. We selected six observations that span over the entire hard state of 2018 outburst, which are highlighted in Fig.~\ref{fig1}. There are three telescopes onboard of \emph{Insight}-HXMT: the high-energy X-ray telescope (HE, 20--250 keV, 5,100 cm$^2$), the medium-energy X-ray telescope (ME, 5--30 keV, 952 cm$^2$), and the low-energy X-ray telescope (LE, 1--15 keV, 384 cm$^2$). There are three types of Field of View (FoV): 1$^\circ$ × 6$^\circ$ (i.e., the small FoV), 6$^\circ$ × 6$^\circ$ (i.e., the large FoV), and the blind FoV that is used to estimate the particle induced instrumental background \cite[see][and references therein]{zhang2020overview}.

The data are processed with hpipeline under \emph{Insight}-HXMT Data Analysis Software (HXMTDAS) version 2.04 \footnote{http://hxmten.ihep.ac.cn/SoftDoc/501.jhtml}. The data are filtered using the criteria recommended by the \emph{Insight}-HXMT team: the pointing offset angle is smaller than 0.04$^\circ$; the elevation angle is larger than 10$^\circ$; the value of the geomagnetic cutoff rigidity is larger than 8; data are used at least 300\ s before and after the South Atlantic Anomaly (SAA) passage. To avoid the possible contamination from the bright earth and nearby sources, only small field of views (FoVs) are applied. Light curves are extracted from screened files using the \textsc{HELCGEN}, \textsc{MELCGEN} and \textsc{LELCGEN} tasks. The lightcurves extracted from 1--10 keV, 10--30 keV and 27--150 keV and the hardness-intensity diagram (HID) with the intensity defined as the total count rate from 1--10 keV and hardness defined as the ratio between the count rates from 3--10 keV and 1--3 keV are plot in Fig.~\ref{fig1}. The lightcurves are not barycentered corrected, considering that it would not affect our results.

\section{HILBERT--HUANG TRANSFORM ANALYSIS} \label{sec3}

The Hilbert--Huang transform (HHT) is a method for analyzing nonlinear and non-stationary signals, which consists of two major steps \citep{huang2008review}: (1) using empirical mode decomposition (EMD) to decompose the signal into a number of independent intrinsic mode functions (IMFs); (2) extracting the instantaneous frequencies and amplitudes of the signal by performing the Hilbert transform of the IMFs. 

\subsection{Empirical Mode Decomposition}

Empirical mode decomposition (EMD) is an iterative sifting process for extracting oscillation modes by subtracting the local means from the original data \citep{huang1998empirical,huang2008review}. 
These oscillatory modes are IMFs. An IMF represents an oscillating wave if it satisfies the following two requirements: (1) in the entire data set, the number of extrema and the number of zero crossings must either be equal or differ at most by one, and (2) at any point, the mean value of the envelope defined by the local maxima and the envelope defined by the local minima is zero. 

The numerical procedure to obtain those IMFs can be concluded with the following steps:

(1) Identify all the local extrema of the data $x(t)$, and form the envelopes defined by the local maxima and minima, respectively, with cubic splines.

(2) Compute the mean values $m_1(t)$ by averaging the upper envelope and lower envelope, and subtract the mean values from the data to get the first component $h_1(t) = x(t)-m_1(t)$.

(3) Test if $h_1(t)$ is an IMF. If the first component is not an IMF, let $h_1(t)$ be the new data set. Continue the steps (1) and (2) until the first component is an IMF.

(4) The first IMF component is called as $c_1(t)$. Let the residual signal $r_1(t) = x(t)-c_1(t)$. Continue the steps (1)--(3) until $r_n(t)$ becomes a monotonic function that no more IMF can be extracted.

Based on the above algorithm, the original signal $x(t)$ can thus be expressed as as the sum of IMFs, and the final residual, $r_n(t)$:

\begin{equation}
    x(t)=\sum_{i=1}^{n} c_i(t)+r_n(t)
    \label{eq1}
\end{equation}

If the original time series contains intermittent processes, the EMD may suffer from the mode mixing problem in which a modulation with the same timescale is distributed across different IMFs \citep{yeh2010complementary}. In this work, we applied a developed modified version of the empirical mode decomposition (EMD), i.e., the fast complementary ensemble empirical mode decomposition (CEEMD), which can reduce the effect of the mode mixing problem \citep{huang2009instantaneous,yeh2010complementary,wang2014computational}. The code we used is from PyEMD (\textsc{v1.21}), an open-source Python package \citep{pyemd}.

\subsection{Hilbert Transform}

The second step of the HHT is the Hilbert transform. After the decomposition step, the IMFs are submitted to this process. For a given data, $x(t)$, the Hilbert transform, $y(t)$, is defined as

\begin{equation}
    y(t)=\frac{1}{\pi } P\int \frac{x(t^{'}) }{t-t^{'}} dt^{'}
    \label{eq2}
\end{equation}

where $P$ is the Cauchy principle value. With this definition, $x(t)$ and $y(t)$ form the complex conjugate pair, so we can have an analytic signal $z(t)$ as

\begin{equation}
    z(t)=x(t)+iy(t)=a(t)e^{i\theta (t)} 
    \label{eq3}
\end{equation}

where time-dependent amplitude $a(t)$ and phase $\theta (t)$ are

\begin{equation}
    a(t)=\sqrt{x(t)^2+y(t)^2} 
    \label{eq4}
\end{equation}
and
\begin{equation}
    \theta (t)=\arctan \frac{y(t)}{x(t)} 
    \label{eq5}
\end{equation}

Therefore, the instantaneous frequency $\omega (t)$ can be defined as

\begin{equation}
    \omega (t)=\frac{\mathrm{d} \theta (t)}{\mathrm{d} t} 
    \label{eq6}
\end{equation}

The instantaneous amplitude $a(t)$ can be defined by the upper envelope of the absolute value of an IMF \citep{huang2009instantaneous}.

\section{Result and Discussion} \label{sec4}

\subsection{Phase Lags of the LFQPO}

Fig.~\ref{fig2} shows a representative example of a 50\ s lightcurve from 27--150 keV band with a $\sim$0.4 Hz LFQPO (ObsID P0114661044). After decomposing the lightcurve we find seven significant IMF components. A $\sim$ 0.4 Hz oscillation is identified as the IMF3. The high-frequency noise (summation from IMF0 to IMF2) and the low-frequency noise (summation from IMF4 to the final residual) are also plotted in Fig.~\ref{fig2}, respectively. Using the adaptive decomposition, these zero-mean oscillatory components can yield physically meaningful instantaneous frequencies by using the Hilbert transform \citep{huang1998empirical,huang2008review}. Since the scope of this work is to discern QPO from noises, further investigation on decomposing each noise component is not our goal here. Therefore, in the following part, we will focus on the study of the IMF3, i.e., the LFQPO. Fig.~\ref{fig3} shows the average Fourier power spectrum of these components.

QPO is known to be energy dependent. One way to study the energy-dependence of the QPO is to track its phase lag. As mentioned above, the phase lag directly measured at the LFQPO frequency from the lag-frequency spectrum is not the intrinsic phase lag. The strong broadband noise would bring an underlying phase-lag continuum that interferes with the measurement of the QPO phase lag \citep{ma2021discovery,zhou2022determination}. Through the HHT method, we obtain the independent light curve of LFQPO (IMF3), which allows us directly measure the QPO intrinsic phase lag without introducing interference from the broad band noise. We decompose the light curves from multiple energy bands and calculate their instantaneous phases through HHT. In order to visualize showing the phase lag of QPO, we plot the superimposed QPO light curves from the 1--2.6 keV and 100--150 keV energy bands, respectively (see Fig.~\ref{fig4} left panel). We find that the QPO shows a strong soft lag, i.e., the soft photons lags the hard ones. 

By calculating the cross-correlation function, we find that there is a soft lag around 0.4 rad between 1--2.6 keV and 100--150 keV (see Fig.~\ref{fig4} right panel). This is opposite to the result given by cross-spectrum in the frequency domain, which shows a hard lag. Fig.~\ref{fig5} shows the QPO phase lag as functions of energy, in which the phase lag directly measured at the LFQPO frequency from the lag-frequency spectrum is referred to as the original phase lag, and the phase lag measured through the HHT method is referred to as the intrinsic phase lag. We can see that the absolute value of the intrinsic soft lag increases with increasing energy, from 0 rad to $\sim$ 0.4 rad. Similar results have also been found in \citet{ma2021discovery}, in which the phase lags were calculated by subtracting the average lag of the band-noise from the original QPO phase lag in the lag-frequency spectrum. We quantitatively compare the intrinsic phase lags calculated by these two methods. As shown in the right panel of Fig.~\ref{fig5}, both methods yield similar trends of the lag-energy relation. However, the absolute soft lags obtained with HHT method are different from those obtained with the method from \citet{ma2021discovery}. There is a possible reason for these differences. The method of \citet{ma2021discovery} is based on the assumption that QPO and noise components are convolutional in the time domain \citep{zhou2022determination}, but EMD is based on the additive relationship between different signals. According to the simulation of \citet{zhou2022determination}, the phase lags are different for the two cases. The relationship between QPO and band-noise is still unclear, but for both the addition and convolution assumptions, similar soft phase lags are observed. This may indicate that the relationship between QPO and noise in the time domain is more complex, such as partially convolutional and partially additive. \citet{ma2021discovery} explained the phase lag behavior of the QPO by employing a compact jet with precession. In this scenario, the high-energy photons come from the bottom part of the jet closer to the black hole, while the precession of the compact jet gives arise to the QPO and allows the high-energy photons to reach the observer first, resulting in a soft lag. This scenario also applies to our results.

\subsection{Modulation of LFQPO}

\begin{figure*}
	\centering\includegraphics[width=\columnwidth]{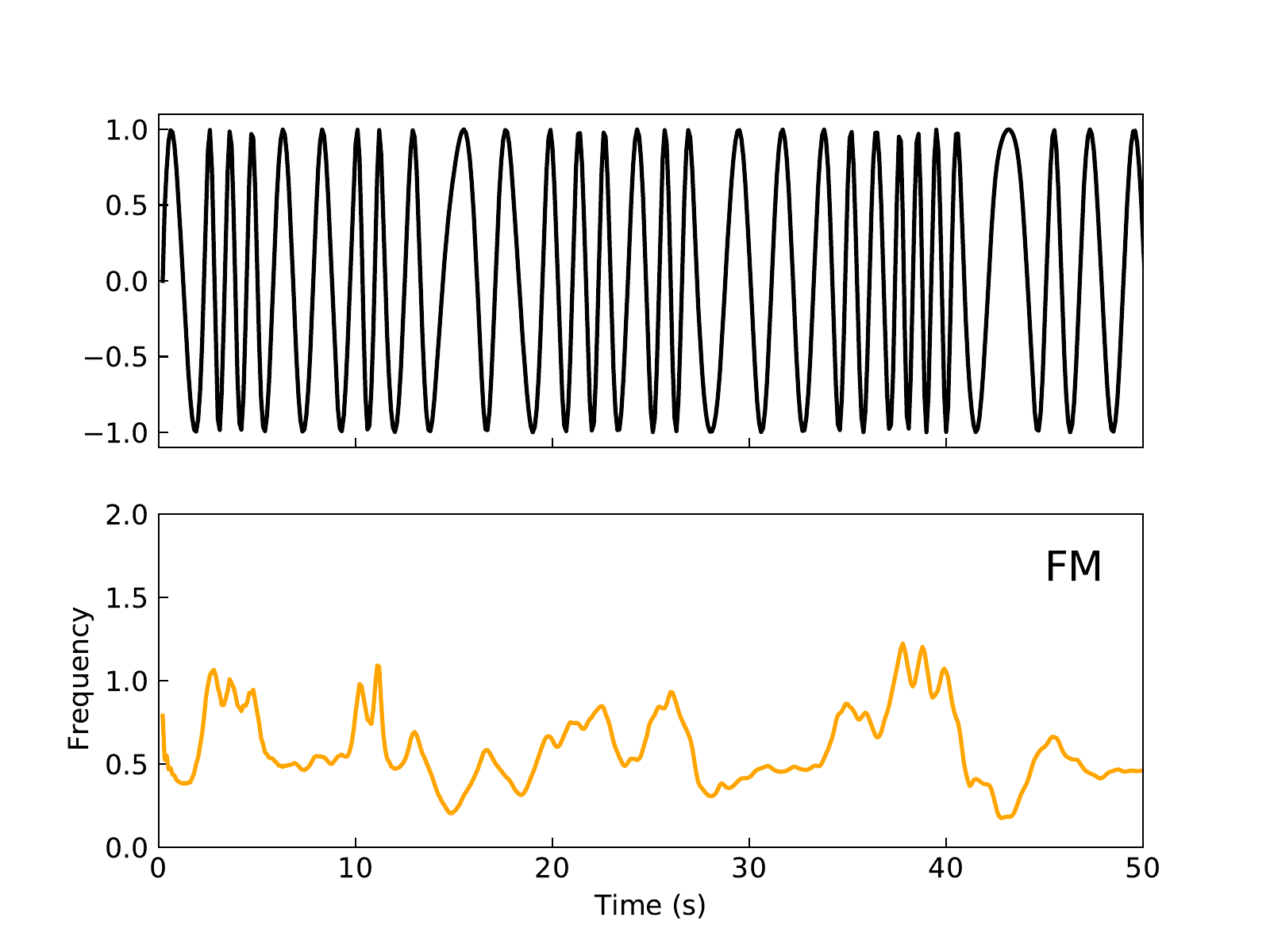}
	\centering\includegraphics[width=\columnwidth]{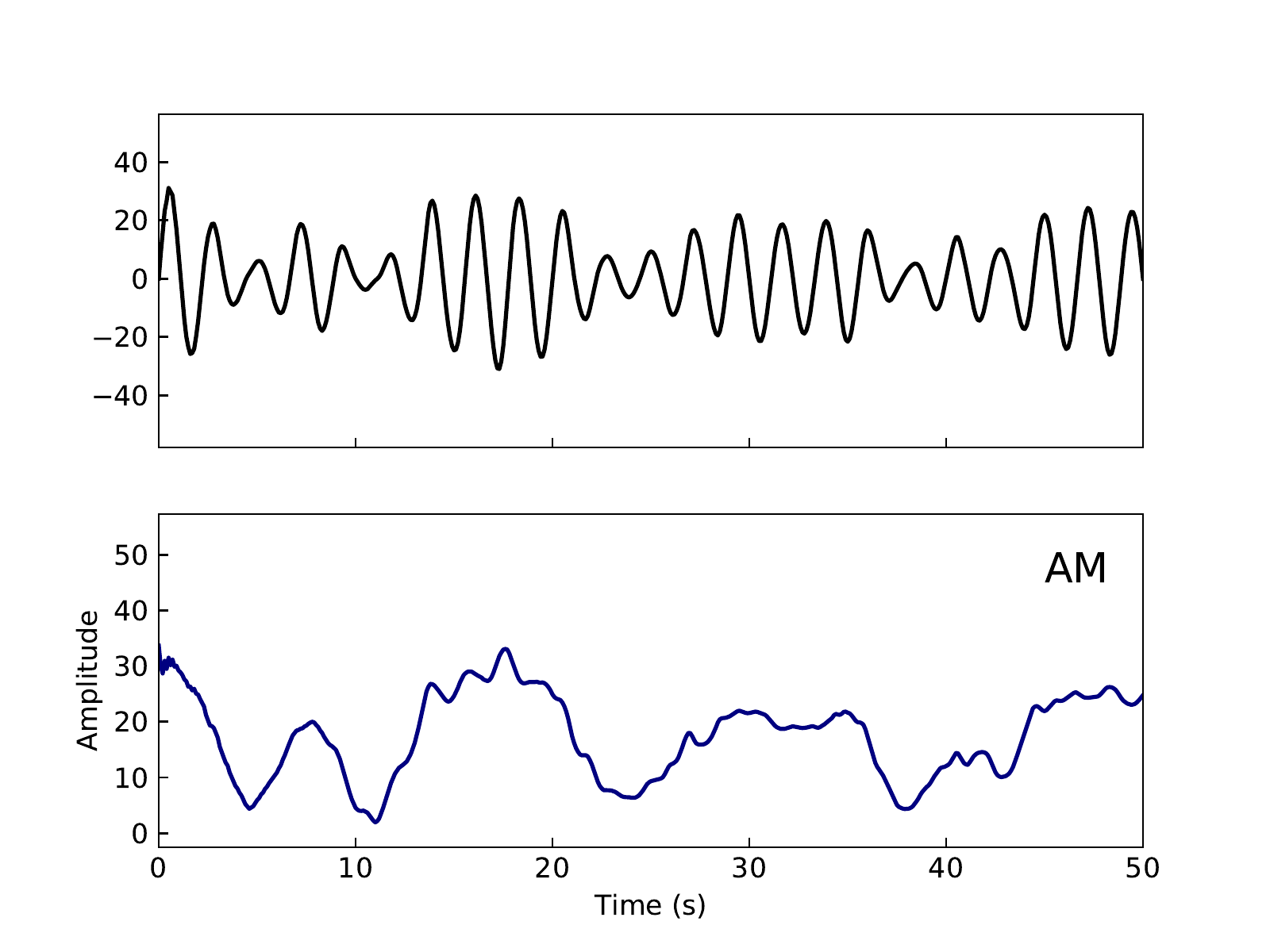}
    \caption{Instantaneous frequency and amplitude of the LFQPO from MAXI J1820+070. The black lines are the simulated light curves.}
    \label{fig7}
\end{figure*}

After measuring the phase lag of the QPO, we move to the second step of the HHT algorithm: using the normalized Hilbert transform \citep{huang2009instantaneous} to extract the instantaneous frequency and amplitude of IMF3. Following \citet{huang2009instantaneous}, we define the instantaneous amplitude of the IMF as the cubic Hermite spline envelope of the local maxima of the absolute values of IMF3. In contrast to the Fourier and wavelet analysis, the calculation of the frequency in HHT is a differentiation over local time domain. Therefore, we can obtain the instantaneous frequency of a signal as long as the sampling time interval, $dt$, is much shorter than the cycle length. This makes it possible for us to explore the origin of the broadening of the peak of LFQPO.

\begin{figure}
	\centering\includegraphics[width=\columnwidth]{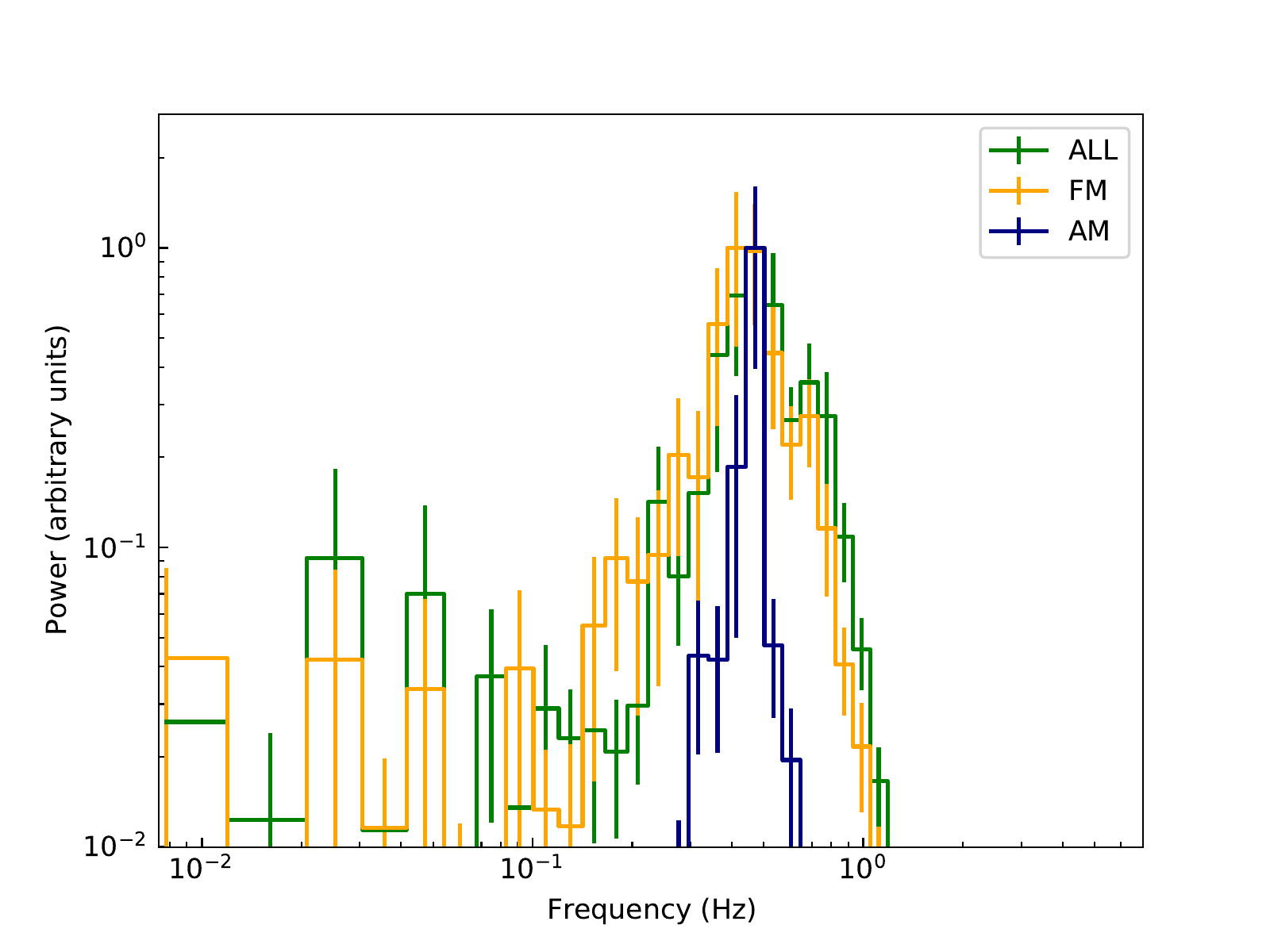}
    \caption{The power density spectra (PDS) of FM (orange), AM (navy) simulated light curves and the original (green) QPO light curve (IMF3). }
    \label{fig8}
\end{figure}

The typical instantaneous frequency and amplitude are shown as the color map for the Hilbert spectrum in Fig.~\ref{fig6}, calculated for the time interval from 0 to 20 s. The color depth represents the magnitude of the amplitude. Fig.~\ref{fig7} shows more detailed instantaneous frequency and instantaneous amplitude information. We see that there are oscillations in both the instantaneous amplitude and the instantaneous frequency of the QPO, which cause the QPO to be quasi-periodic. But with the current information, we cannot tell which kind of the oscillation is dominant. To explore this, we generate two simulated light curves using the instantaneous amplitude and frequency information (Fig.~\ref{fig7}). One of the light curves only includes the frequency modulation, leaving the amplitude be constant. The other light curve only includes the amplitude modulation, leaving the frequency remain at the center frequency of the QPO. Next, we perform Fourier transform on these two light curves to produce their power density spectra (PDS), as shown in Fig.~\ref{fig8}. For better comparison, the figure also includes the PDS generated from the original QPO light curve. As shown in the figure, the effect of frequency modulation (orange) on QPO broadening is significantly stronger than that of amplitude modulation (navy). This suggests that frequency modulation plays a major role in the broadening of the peak of LFQPO. 

According to the simulation from \citet{ingram2019review}, if the quasi-period of QPO is dominated by frequency modulation, then its fundamental frequency and harmonic frequency of QPO should have the same Q factor. On the contrary, if it is dominated by amplitude modulation, then the fundamental and harmonic frequencies of the QPO should have the same width. By fitting the original power spectrum, we find that the Q-factors of the QPO fundamental and second harmonic are $5.5 \pm 0.1$ and $5.7 \pm 0.2$, suggesting a frequency modulation. Combined these results with those we get from the HHT, we conclude that the quasi-periodic of QPO in MAXI J1820+070 is mainly caused by the frequency modulation. 

\subsection{Origin Of Modulation}

\begin{figure}
	\centering\includegraphics[width=\columnwidth]{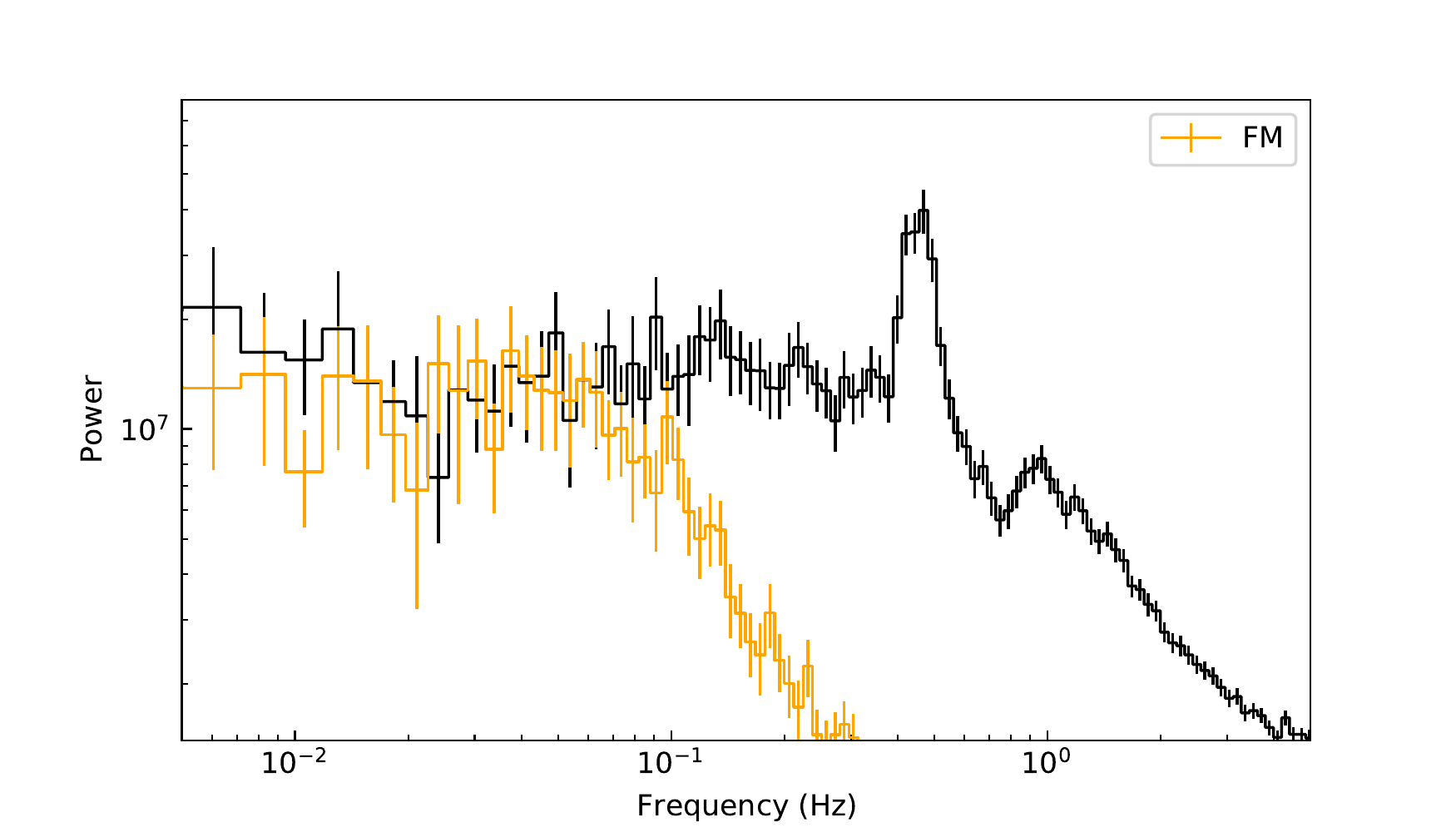}
	\centering\includegraphics[width=\columnwidth]{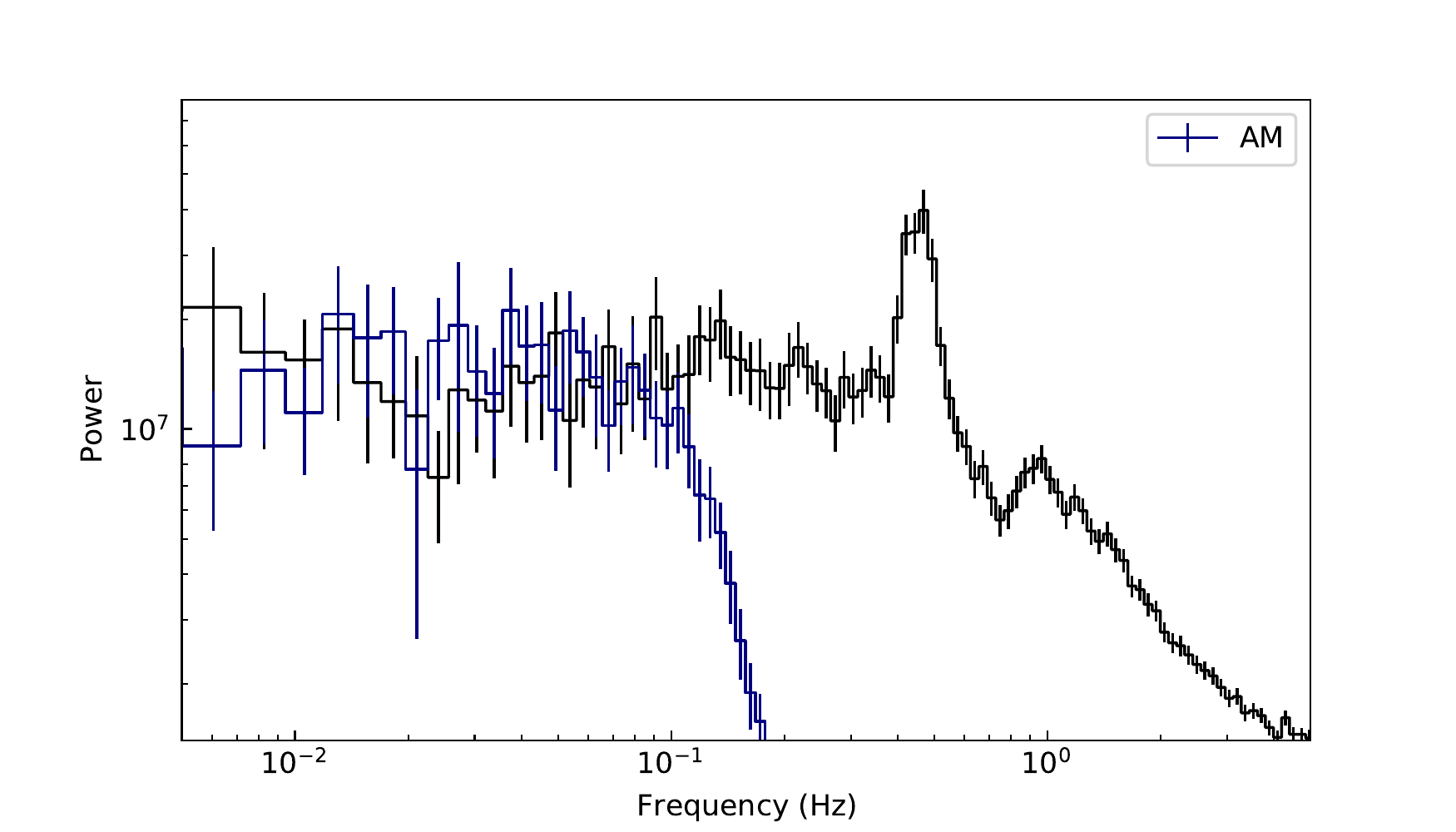}
    \caption{The power density spectra (PDS) of the instantaneous frequency and instantaneous amplitude curves for IMF3.}
    \label{fig9}
\end{figure}

\begin{figure*}
	\centering\includegraphics[width=\columnwidth]{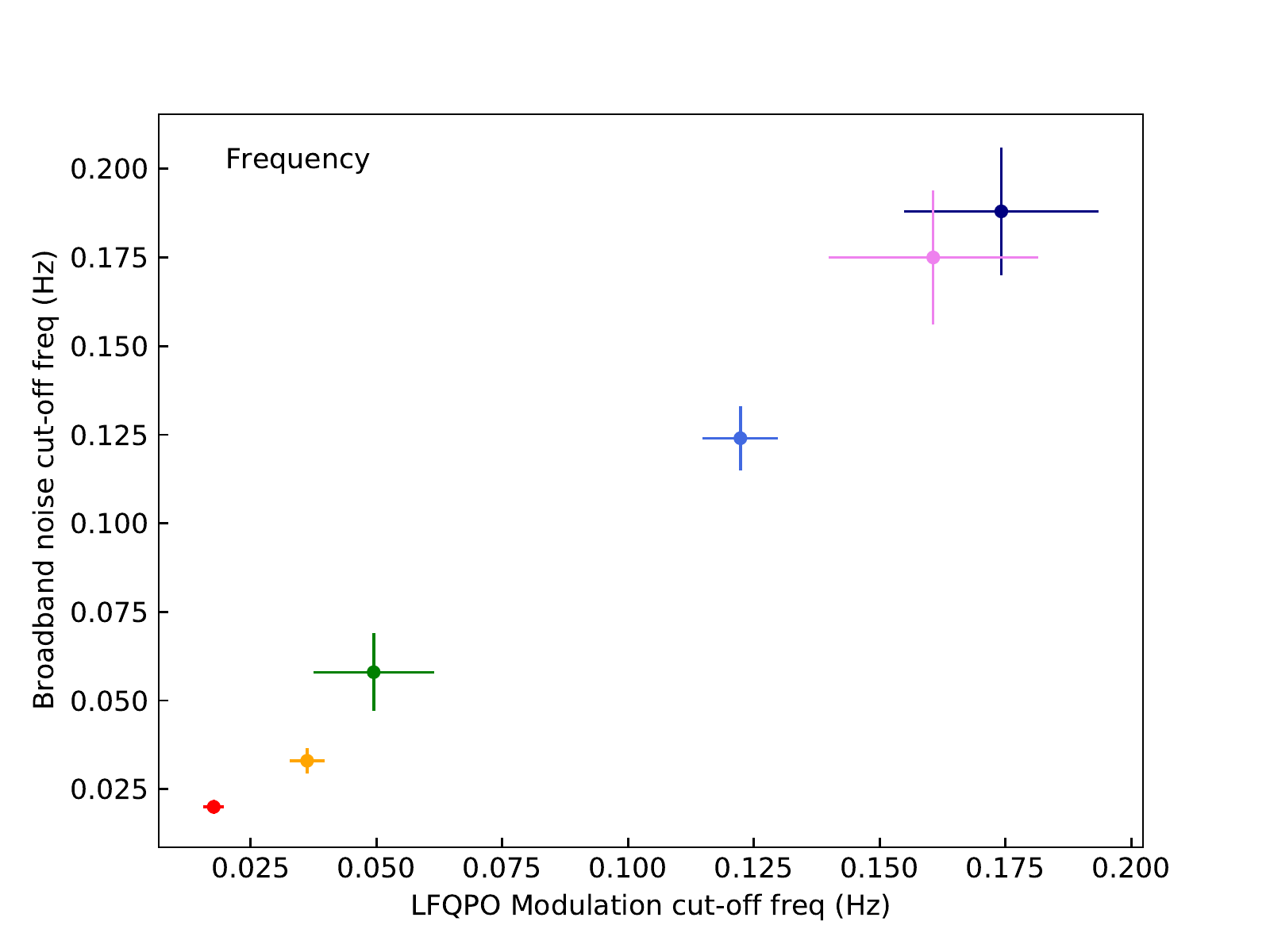}
	\centering\includegraphics[width=\columnwidth]{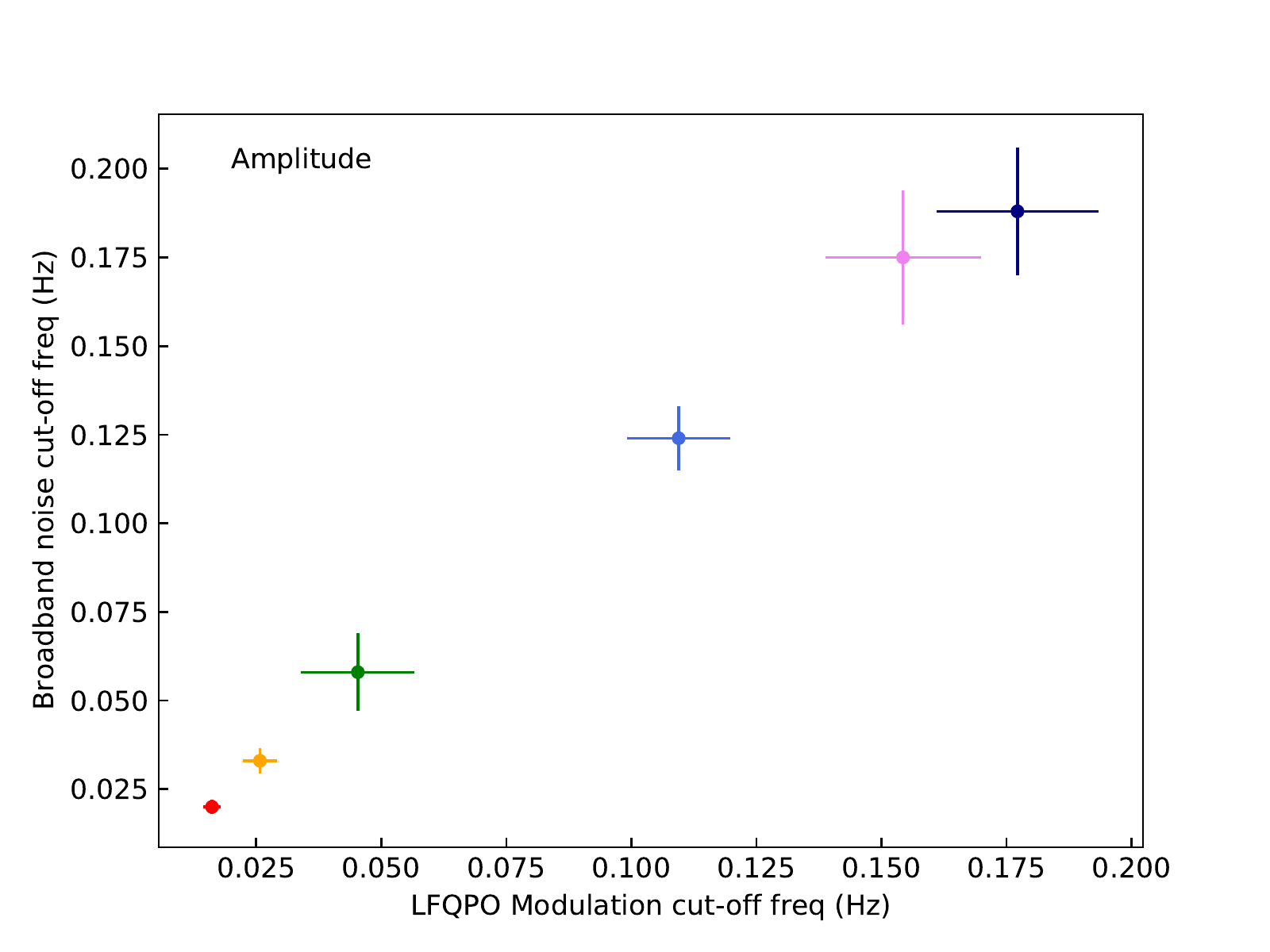}
    \caption{The cut-off frequency of broadband-noise as a function of that of amplitude modulation (left panel) or frequency modulation (right panel). The colour of each observation matches that in Fig.~\ref{fig1}.}
    \label{fig10}
\end{figure*}

To explore the origin of the modulation, we calculate the power spectra of instantaneous frequency and instantaneous amplitude, respectively (see Fig.~\ref{fig9}).  
We find that both them show a shape similar to that of broad-band noise, implying that the modulation is not caused by a completely random process. To verify this, we decompose more observation light curves from the hard state, which are colored marked in Fig.~\ref{fig1}. By fitting the power spectrum with multiple Lorentzian functions, we get the cut-off frequency for each component. As shown in Fig.~\ref{fig10}, there is a strong linear correlation between the timescale of the frequency/amplitude modulation and the broad-band noise. This may suggest that the modulation of QPO may have a common origin with broad-band noise. 

To further investigate what leads to the modulation, we first need to understand how the QPO originates. The origin of QPO has been studied for a long time, and several models have been proposed. One of the most accepted models is the Lense-Thirring (L-T) precession model, which assumes that QPO is generated by the relativistic precession of an inner accretion flow \citep{ingram2009low,you2018x,you2020x} or a jet \citep{ma2021discovery}. 

For the first case, the modulation may come from the fluctuation propagation in the hot accretion flow. The inward propagation of the fluctuations causes changes in the surface density of the accretion flow, thus altering the hot flow’s moment of inertia and inducing the precession frequency jitters. Besides, the fluctuations propagation model \citep{lyubarskii1997flicker,ingram2012modelling,ingram2016modelling,mushtukov2019broad} is also widely used to explain the broad-band noise in the power spectrum \citep{rapisarda2014evolution,rapisarda2016testing,ingram2016modelling,rapisarda2017cross,rapisarda2017modelling,turner2021investigating,yang2022accretion}. Thus, it can naturally explain why the modulation has a shape similar to that of the broad-band noise in the PDS. However, this explanation has its shortcomings. Fluctuations propagation can generate not only low-frequency noise, but also high-frequency noise. We observed both low-frequency and high-frequency noises in the power spectrum, but only low-frequency modulation was observed. This cannot be well explained under the fluctuations propagation model. 

As for the second case, it assumes that the QPO is generated by the precession of a jet. The modulation can be explained by the internal shock model of 
the jet \citep{rees1978m87,spada2001internal,boettcher2010timing}. In this model, shells of gas are continuously ejected with randomly variable velocities and then propagate along the jet. At some points, the fastest fluctuations start catching up and merging with slower ones. This leads to the formation of shocks in which electrons are accelerated up to relativistic energies. \citet{malzac2014spectral} have shown that internal shocks caused by fluctuations of the outflow velocity can produce shapes on the power spectrum that resemble low-frequency broad-band noise. The jet behaves like a low-pass filter, as the shells of plasma colliding and merging with each other, the highest frequency velocity fluctuations are gradually damped and the size of the emitting regions increases \citep{malzac2014spectral}. This model can well explain the absence of high frequency modulation. Moreover, the jet precession model can also naturally explain the large soft phase lags of QPO \citep{ma2021discovery}. 

\renewcommand\arraystretch{1.2}

\section{CONCLUSION} \label{sec5}

In this paper, we performed the HHT analysis of the LFQPOs in MAXI J1820+070. With the EMD method, we are able to extract the independent light curve of the QPO and measure the QPO intrinsic phase lag. We find a soft QPO phase lag in this source (low energy photons lag behind high energy photons), and the absolute value of the QPO phase lag increases with energy. Our result is different from the phase lag calculated from the lag-frequency spectrum, in which the lag includes the contribution from the broadband noises. Our results show that the EMD method can significant reduce the interference from the broad band noise on the measurement of the QPO intrinsic lag. 

By analyzing the instantaneous frequency and instantaneous amplitude obtained from HHT, we find that the broadening QPO peak in the power spectrum of MAXI J1820+070 is dominated by the frequency modulation. Through further analysis, we find that this modulation probably share a common physical origin with the broad-band noise, and can be well explained by the internal shock model of the jet.

\begin{acknowledgments}
This work has made use of the data from the \textit{Insight}-HXMT mission, a project funded by China National Space Administration (CNSA) and the Chinese Academy of Sciences (CAS), and data and/or software provided by the High Energy Astrophysics Science Archive Research Center (HEASARC), a service of the Astrophysics Science Division at NASA/GSFC. This work is supported by the National Key RD Program of China (2021YFA0718500) and the National Natural Science Foundation of China (NSFC) under grants U1838201, U1838202, 11733009, 11673023, U1938102, U2038104, U2031205, the CAS Pioneer Hundred Talent Program (grant No. Y8291130K2) and the Scientific and Technological innovation project of IHEP (grant No. Y7515570U1). This work was partially supported by International Partnership Program of Chinese Academy of Sciences (Grant No.113111KYSB20190020).
\end{acknowledgments}

%



\appendix
\setcounter{table}{0}
\setcounter{figure}{0}
\renewcommand{\thetable}{A\arabic{table}}
\renewcommand{\thefigure}{A\arabic{figure}}

\begin{figure}[htbp]
\centering
\subfigure{
\includegraphics[width=8.5cm]{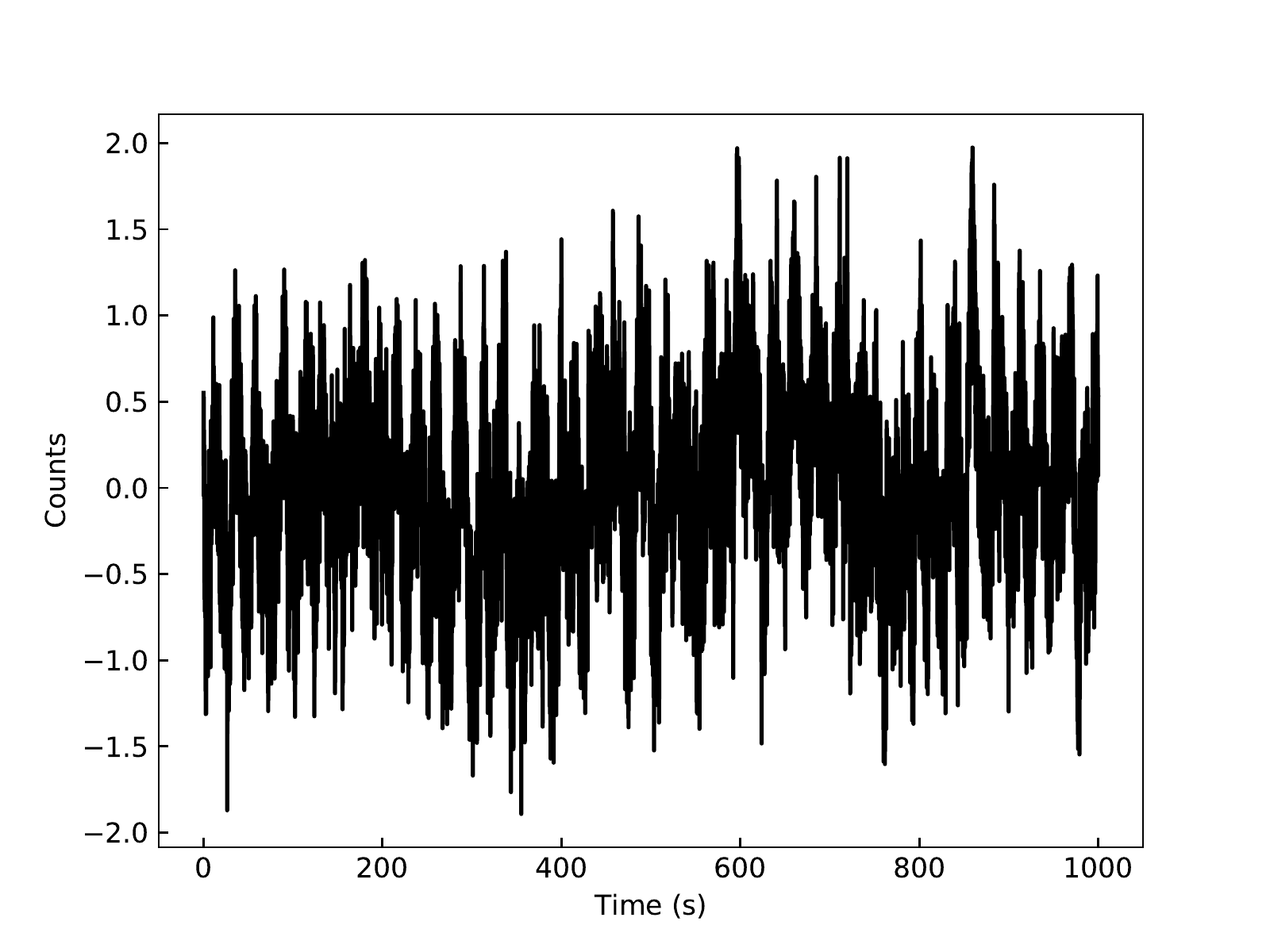}
}
\quad
\subfigure{
\includegraphics[width=8.5cm]{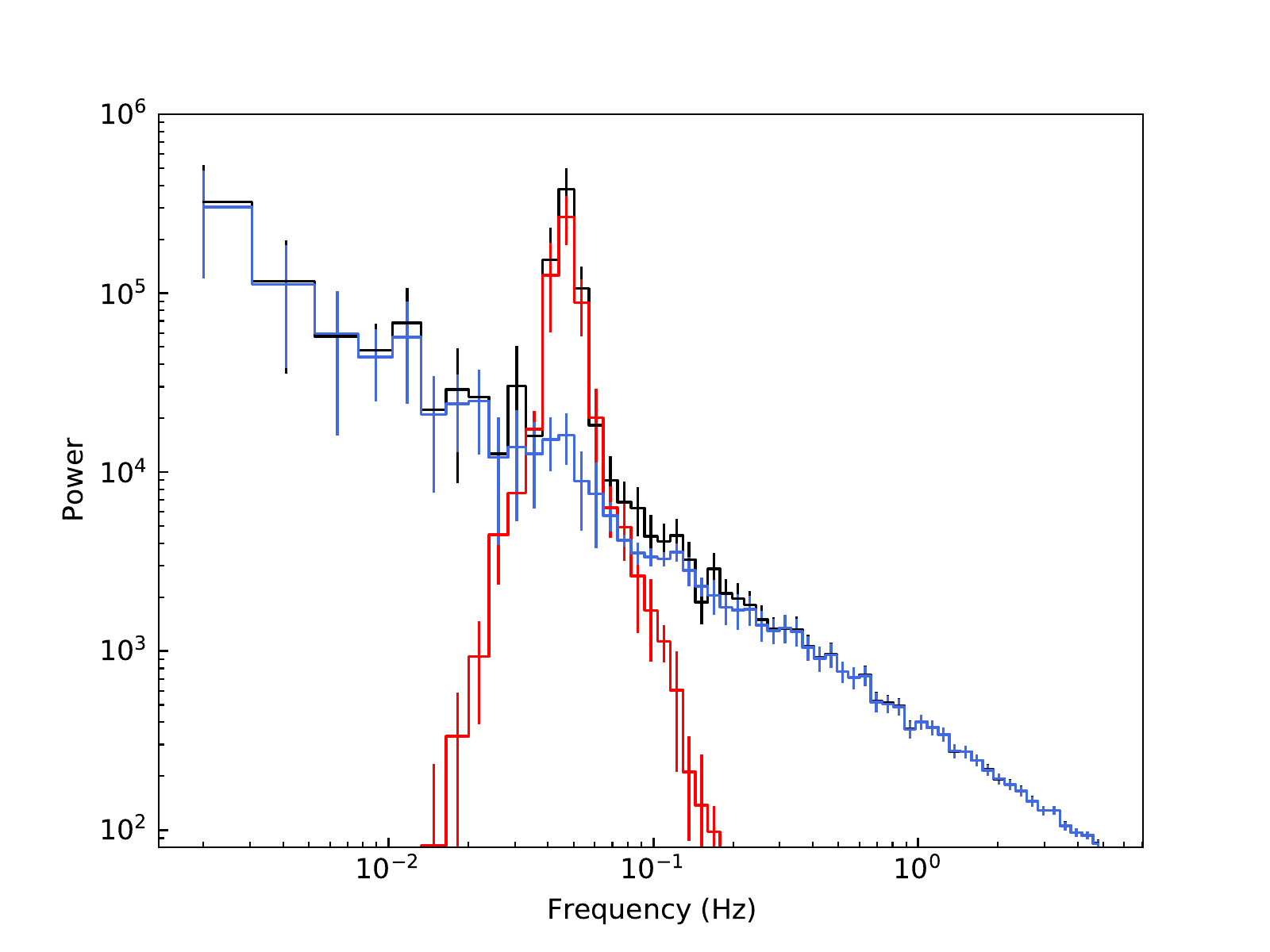}
}
\caption{Left: Simulated light curve composed of red noise and QPO. Right: Fourier power spectra that are produced from the original light curve (black), decomposed QPO component (red) and decomposed noise component (blue), separately. }
\label{fig11}
\end{figure}

\begin{figure}[htbp]
\centering
\subfigure{
\includegraphics[width=13cm]{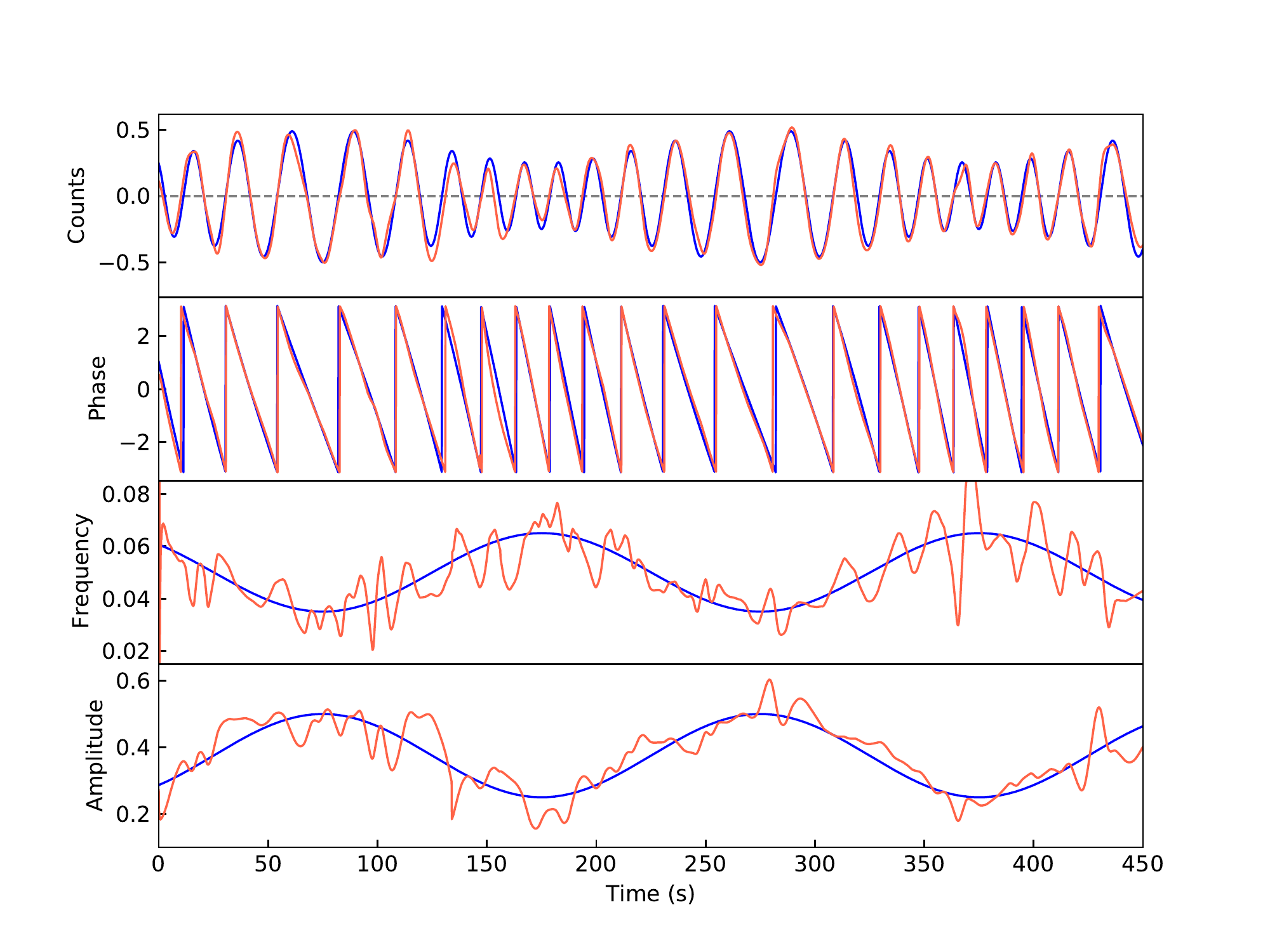}
}
\caption{The QPO lightcurves and their Hilbert spectra. The blue lines represent the original QPO component while the red lines represent the decomposed QPO component.  }
\label{fig12}
\end{figure}

In order to examine the credibility of the results of QPO analysis, we perform the robustness tests on the HHT method. First, we simulated a red noise light curve using the Timmer-Koenig method \citep{timmer1995generating}, then added a sinusoidal signal with frequency and amplitude modulation as the QPO component. And then we decomposed the synthetic light curve using the CEEMD method. The results are shown in Fig.~\ref{fig11}, in which the red line represents the QPO component and the blue line represents the noise component. Our results show that the CEEMD algorithm can effectively separate the QPO and noise components. Subsequently, we compared the decomposed QPO with the original QPO light curves and found that the QPO profile was accurately recovered (see Fig.~\ref{fig12}). Further Hilbert transform analysis on the decomposed QPO revealed that the instantaneous phase, frequency, and amplitude trends of the decomposed QPO are highly consistent with those of the original signal, but with local fluctuations, which is possibly due to the mode mixing between the QPO and noise components.

\begin{figure}[htbp]
\centering
\subfigure{
\includegraphics[width=8.5cm]{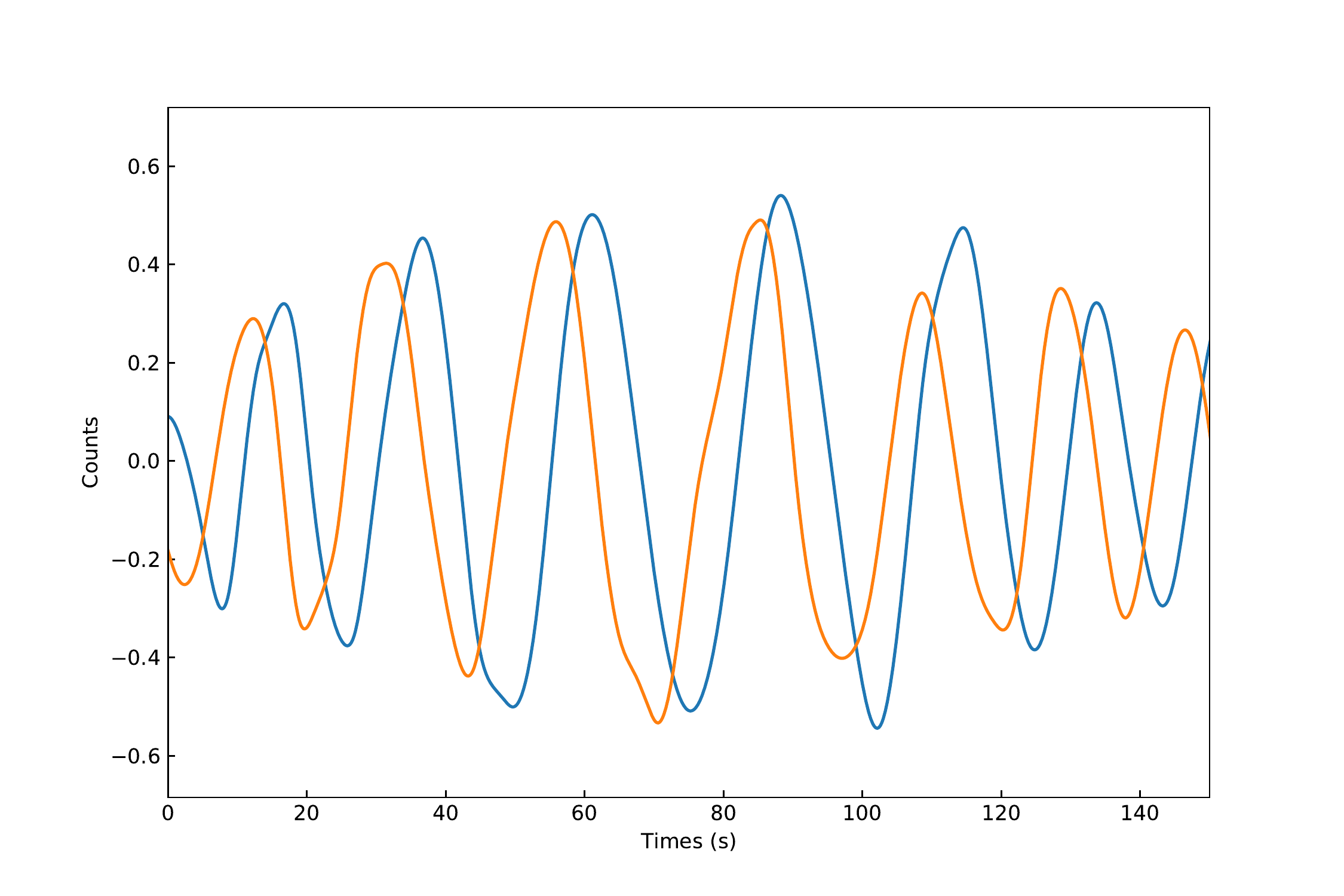}
}
\quad
\subfigure{
\includegraphics[width=8.5cm]{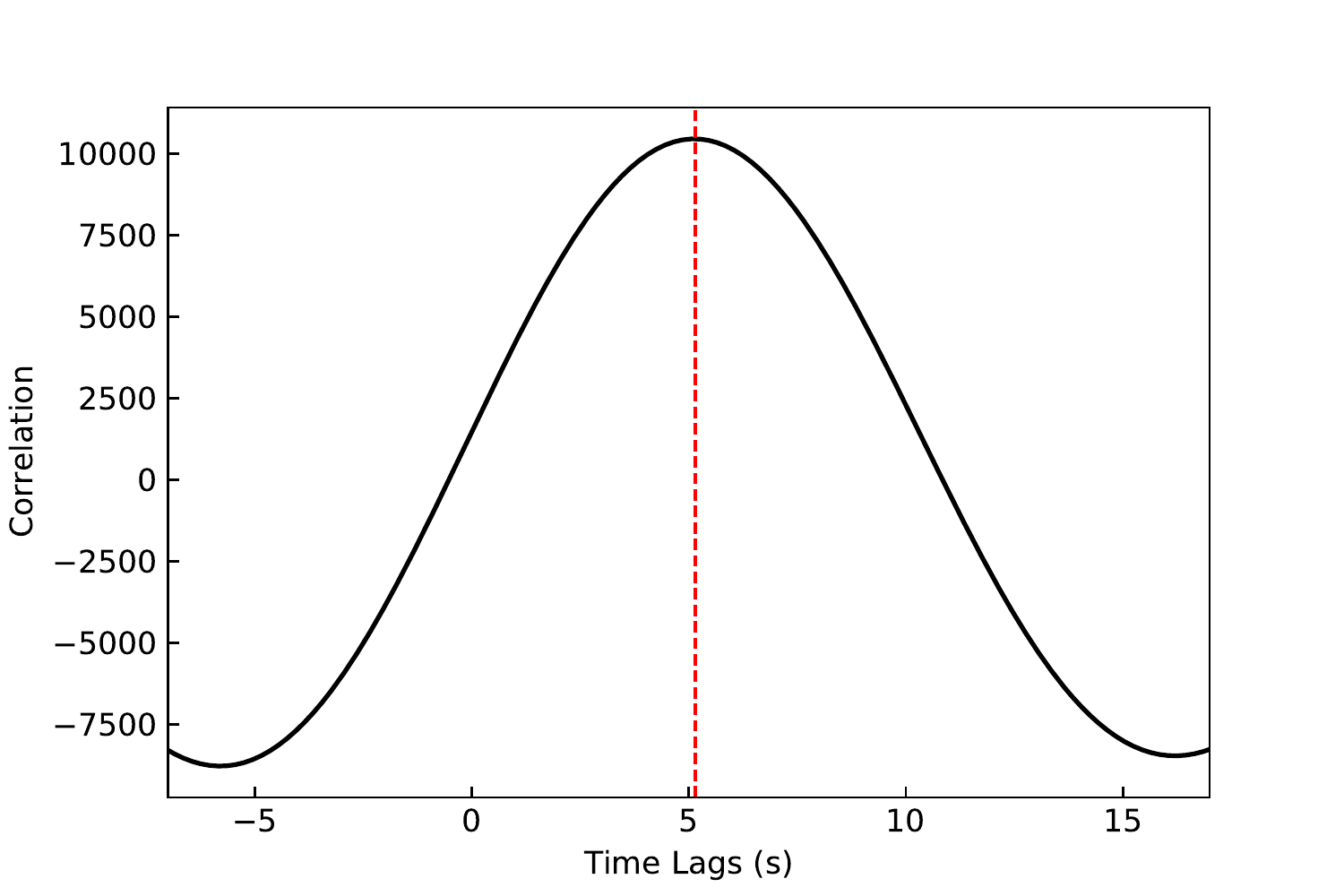}
}
\caption{The lightcurves and cross-correlation function of the decomposed QPO components. }
\label{fig13}
\end{figure}

To investigate whether such fluctuations affect the measurements of the intrinsic QPO phase lag, we simulated another light curve with a $+5s$ time lag (equivalent to $+\pi/2$ phase lag) into the QPO component and a $-10s$ time lag into the noise component. We then decomposed this light curve and measured the time lags between the two decomposed  QPO light curves, which is shown in Fig.~\ref{fig13}. The two decomposed QPO components exhibited a clear positive time lag very close to $5s$, indicating that the HHT method works very well to measure the intrinsic phase lag of QPOs.

\bibliography{sample631}{}
\bibliographystyle{aasjournal}



\end{document}